\documentclass[journal,12pt,draftclsnofoot,onecolumn]{IEEEtran}
\IEEEoverridecommandlockouts
\usepackage{verbatim}
\usepackage{cite}
\newcommand{\upcite}[1]{\textsuperscript{\!\!\cite{#1}}}%
\usepackage{amsmath,amssymb,amsfonts}
\usepackage{graphicx}
\usepackage{textcomp}
\usepackage{xcolor}
\usepackage{epsfig}
\newtheorem{Remark}{\it Remark}[section]

\newtheorem{Proposition}{\it Proposition}[section]
\newtheorem{Lemma}{\it Lemma}[section]
\newtheorem{Corollary}{\it Corollary}[section]

\makeatletter
\newcommand{\Rmnum}[1]{\expandafter\@slowromancap\romannumeral #1@}
\makeatother
\usepackage{bm}
\usepackage{multicol}
\usepackage{setspace}
\usepackage{subfigure}
\usepackage{epstopdf}
\usepackage{fancyhdr} 
\usepackage{indentfirst}
\usepackage{enumerate}
\usepackage{stfloats}
\usepackage{bm}
\usepackage{multirow}
\usepackage{threeparttable}
\usepackage{CJK}
\usepackage[justification=centering]{caption}
\usepackage{makecell}
\usepackage{caption}
\usepackage{cases}
\usepackage{amsmath}
\usepackage{algorithm}
\usepackage{algpseudocode}

\usepackage{graphics}
\usepackage{caption}
\usepackage{booktabs}

\definecolor{deepblue}{rgb}{0.6,0,0.4}
\definecolor{red}{rgb}{0.8,0,0}
\definecolor{blue}{rgb}{0.2,0,0.8}

\def\BibTeX{{\rm B\kern-.05em{\sc i\kern-.025em b}\kern-.08em
		T\kern-.1667em\lower.7ex\hbox{E}\kern-.125emX}}
\begin{document}
 \title{Fundamental Limitation of Semantic Communications: Neural Estimation for Rate-Distortion}
 \author{\IEEEauthorblockN{Dongxu Li, Jianhao Huang,  Chuan Huang, Xiaoqi Qin, Han Zhang, and Ping Zhang}
   \thanks{
      Manuscript accepted to Journal of Communications and Information Networks November 15, 2023.  Part of this work was presented in  IEEE GLOBECOM 2023\upcite{li_confer}. (Corresponding author: Chuan Huang.)
      
D. Li and C. Huang are with the School of Science and Engineering and the Future Network of Intelligence Institute, the Chinese University of Hong Kong, Shenzhen, 518172 China. Emails: dongxuli@link.cuhk.edu.cn, huangchuan@cuhk.edu.cn.

J. Huang is with the Department of Electrical and Electronic Engineering, the University of Hong Kong, Hong Kong 999077, China. Emails: Jianhaoh@hku.hk. 

X. Qin, H. Zhang, and P. Zhang are with the School of Information and Communication Engineering and the State Key Laboratory of Networking and Switching Technology,  Beijing University of Posts and Telecommunications, Beijing 100876, China. Email: xiaoqiqin@bupt.edu.cn, hanzh92@bupt.edu.cn, and  pzhang@bupt.edu.cn.} 
}

\maketitle
 
\begin{abstract}
This paper studies the fundamental limit of semantic communications over the discrete memoryless channel.  We consider the scenario to send a semantic source consisting of an observation state and its corresponding semantic state, both of which are recovered at the receiver. To derive the performance limitation, we adopt the semantic rate-distortion function (SRDF) to study the relationship among the minimum compression rate, observation distortion, semantic distortion, and channel capacity. For the case with unknown semantic source distribution, while only a set of the source samples is available, we propose a neural-network-based method by leveraging the generative networks to learn the semantic source distribution. Furthermore, for a special case where the semantic state is a deterministic function of the observation, we design a cascade neural network to estimate the SRDF.  For the case with perfectly known semantic source distribution, we propose a general Blahut-Arimoto algorithm to effectively compute the SRDF. Finally, experimental results validate our proposed algorithms for the scenarios with ideal Gaussian semantic source and some practical datasets. 
   
	\end{abstract}
	\section{Introduction}

With the extensive deployment of artificial intelligence in wireless communications, semantic communication is emerging as a hot research area for future communication systems\upcite{gunduz2022beyond, zhang2022toward, strinati20216g}. Unlike conventional communication technologies that aim to accurately transmit bit information from the transmitter to the receiver, semantic communication is geared towards some specific semantic tasks, e.g., security monitoring\upcite{9398576} and edge inference\upcite{9606667}. Moreover, by extracting and encoding desired information from the source data that is most relevant to the considered semantic tasks, semantic communication dramatically reduces resource consumption for communications. Therefore,  semantic communication is expected to flourish in many practical scenarios, e.g., virtual reality\upcite{shi2023task} and smart cities\upcite{9679803}.

 Semantic communication was first introduced by Shannon\upcite{Shannon} and Weaver\upcite{weaver} in the 1950s, aiming to study the precise transmissions of semantic information. Since then, researchers have been working for several decades on how to define and model semantic information. Carnap and Bar-Hillel\upcite{Carnap} proposed a logical probability-based approach to replace the statistical one in classical information theory where the amount of information is determined by its statistical rarity. Floridi\upcite{floridi2004outline} further proposed a semantic information theory based on logical probabilities, aiming to resolve the semantic paradox problem\upcite{sequoiah2007metaphilosophy}.  However, the subjectivity of semantic understanding among different people poses a challenge in designing the logical probability functions that can be widely applied in practice. Moreover, many recent works focused on deep learning (DL)-based joint source-channel coding (JSCC)\upcite{8723589,9066966,9791398}, i.e., the source coding and channel coding are jointly optimized by using deep neural networks, and DL-based separate source-channel coding (SSCC)\upcite{huang_confer,9191247,huang2023joint}, i.e., a deep neural network is designed to compress source symbols, followed by a classical channel coding scheme, e.g., low-density parity check coding\upcite{gallager1968information}, to implement the point-to-point semantic communications. 
 
   Recently, a new type of semantic source model was discussed in Ref. \cite{liu2021rate} and \cite{9844779}, and the semantic source is modeled as two distinct parts: an extrinsic observation and an intrinsic semantic state. Taking video as an example, the video signal itself
 represents the extrinsic observation, and its features, which are generated based on this video for certain tasks, such as action recognition\upcite{pareek2021survey} and object detection\upcite{Yang}, correspond to the intrinsic semantic state. Two distortion measures, namely the observation and semantic distortions, were adopted to study the semantic rate-distortion function (SRDF)\upcite{liu2021rate,9844779}, which is defined as the minimum compression rate subject to the maximum tolerable observation and semantic distortions. In this semantic source model, semantic state is generally not observable while can be inferred from the observation\upcite{liu2021rate}. Then, the source encoder only focuses on encoding the observation and the decoder reconstructs both
the semantic state and the observation, subject to both the semantic and observation distortion constraints. Under this framework, it has been shown\upcite{9844779,Thomas} that SSCC performs the same as  JSCC over the infinite discrete memoryless channels.
However, SRDF generally has no closed-form expression\upcite{9844779}, and can only be effectively solved for some specifically distributed semantic sources, e.g., Gaussian\upcite{9844779} and binary sources\upcite{9834593}.  Considering the high-dimensional feature of the semantic sources, e.g., text\upcite{Qin2}, speech\upcite{Qin}, and images\upcite{9959884}, their distributions are general and difficult to be well modeled, making it extremely challenging to compute the corresponding SRDFs.

Despite many efforts already made in the field of semantic communications, there remains an absence of theoretical research on the fundamental analysis of SRDFs for generally distributed sources. To deal with this issue, the purpose of this paper aims to analyze the fundamental limits of the point-to-point semantic communications for general semantic sources. Specifically, inspired by the semantic source model proposed in Ref. \cite{liu2021rate}, we consider a semantic source pair at the transmitter consisting of the extrinsic observation and its intrinsic semantic state. Here, only the observation data is compressed and encoded at the transmitter and then sent over a discrete memoryless channel to the receiver, where both the extrinsic observation and the semantic state are finally recovered. We aim to compute the corresponding SRDF for generally distributed semantic sources,  which reveals the trade-off among the compression rate, observation distortion, semantic distortion, and channel capacity. Moreover, for the case with imperfectly known semantic source distribution, i.e., only a certain amount of source samples are available, motivated by the neural estimation method for traditional rate-distortion functions\upcite{10124059}, we design a neural estimator for SRDF: first, we show that SRDF can be rewritten as an inf-sup problem via its dual property; then, we leverage generative networks to solve this problem and derive the neural estimation of the SRDF (NESRD), and further show NESRD to be a strongly consistent estimator; finally,  when the semantic state is a deterministic function of the observation, we design a cascade neural network framework to train the derived NESRD.  For the case with perfectly known semantic source distribution, we generalize the conventional Blahut-Arimoto (BA) algorithm to numerically compute SRDF and analyze the computational complexity of this proposed algorithm.
  
  The rest of this paper is organized as follows. Section II introduces the SSCC-based framework. 
  Section III derives the neural estimator for SRDF under the case with unknown semantic source distributions.  Section IV proposes the general BA algorithm to calculate SRDF for the case with known semantic source distributions.    Experimental results are presented and discussed in Section V to validate the theoretical results. Finally, Section VI concludes this paper. 
  
  \emph{Notation}: $ \log(x) $ and $ \ln(x) $ denote base-2 and natural logarithms, respectively;$e^{x} $ stands for natural exponent; $|\mathcal{X}| $ denotes the size of  finite alphabet $ \mathcal{X} $; $\max\{x, y\}$ is the maximum value between two real numbers $ x $ and $ y $; $ \textbf{E}_{P_X}(\cdot) $ is the expectation for random variable $ X $ with probability distribution $ P_X $.

   \section{System Model}   
  We consider a general semantic communication system, where the transmitter compresses the source data and then transmits it through a point-to-point discrete memoryless channel to the receiver for data recovery and processing specific semantic tasks.  Here, a memoryless semantic source is modeled as a pair of random variables\footnote{$X$ and $S $ can be random vectors for the case that they are obtained from high-dimensional sources, e.g., text\upcite{Qin2} and images\upcite{9959884}.} $ (X, S) $ with a joint probability distribution $ (X, S) \sim P_{(X,S)}  $ supported on a finite product alphabet  $ \mathcal{X}\times \mathcal{S} $, where $ X $ represents the extrinsic observation of the source, $ S $ is the intrinsic semantic state relevant to the considered semantic task, and $ \mathcal{X}$ and $ \mathcal{S}$ are the alphabets of $ X$ and $S $, respectively. 
  
    \begin{figure}[htbp]
	\centering
	\includegraphics[width=4
	in]{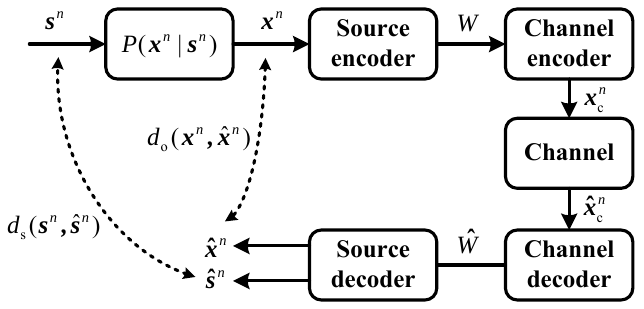}
	\caption{ Framework of  SSCC scheme for semantic communications.}
	\label{fig_system} 
    \end{figure}
  
    As shown in Fig. \ref{fig_system}, an SSCC-based semantic communication framework over the discrete memoryless channel is considered in this paper. The transmitter sends a sequence of length-$n$ independent and identically distributed (i.i.d.) samples of source pair $ (X,S) $, denoted as $(\boldsymbol{x}^n,\boldsymbol{s}^n)=(x_1,\cdots,x_n,s_1,\cdots,s_n)$. 
      The source encoder has access to observation sequence $ \boldsymbol{x}^n $, whereas $\boldsymbol{x}^n$ is related to semantic state sequence $ \boldsymbol{s}^n $ through a conditional probability $ P(\boldsymbol{X}^n=\boldsymbol{x}^n|\boldsymbol{S}^n = \boldsymbol{s}^n) $. Then, $ \boldsymbol{x}^n $ is compressed by the source encoder  into an index $ W \in \{1, \cdots,2^{nR_{\text{s}}}  \} $, with $ R_{\text{s}} $ being the compression rate. After that, $ W $ is encoded by the channel encoder into a length-$n$ channel codeword $\boldsymbol{x}_{\text{c}}^n $.
       Then, $ \boldsymbol{x}_{\text{c}}^n $ is transmitted via the discrete memoryless channel, and $\hat{\boldsymbol{x}}_{\text{c}}^n $ is the received symbol sequence. At the receiver, the channel decoder decodes $\hat{\boldsymbol{x}}_\text{c}^n $ as an estimated index $ \hat{W} $. Finally,  the source decoder recovers the observation and semantic state sequences as $ \hat{\boldsymbol{x}}^n =(\hat{x}_1,\cdots,\hat{x}_n) $ and $ \hat{\boldsymbol{s}}^n =(\hat{s}_1,\cdots,\hat{s}_n) $, respectively.  
       
       We denote reproduction observation and semantic states as $ \hat{X} $ and $ \hat{S} $, respectively, taking values from alphabets $ \mathcal{\hat{X}} $ and $ \mathcal{\hat{S}} $, respectively. Then, we define $ d_{\text{o}}: \mathcal{X}\times\mathcal{\hat{X}}\to [0, + \infty)$ as the single-letter observation distortion between observation state $ X $ and its reproduction $ \hat{X} $,  and define $ d_{\text{s}}: \mathcal{S}\times\mathcal{\hat{S}}\to [0, + \infty)$ as the single-letter semantic distortion between semantic state $ S $  and its reproduction $\hat{S} $. Correspondingly, the block-wise observation and semantic distortion measures are defined as $ d_{\text{o}}(\boldsymbol{x}^n,\hat{\boldsymbol{x}}^n) \triangleq \frac{1}{n}\sum\limits_{i=1}^nd_{\text{o}}(x_i, \hat{x}_i ) $ and $ d_{\text{s}}(\boldsymbol{s}^n,\hat{\boldsymbol{s}}^n)\triangleq \frac{1}{n}\sum\limits_{i=1}^nd_{\text{s}}(s_i, \hat{s}_i ) $, respectively\upcite{9844779}.
     Then, denote $ D_{\text{o}} $  and $ D_{\text{s}} $ as the maximum tolerable observation and semantic distortions, respectively, and characterize the minimum of compression rate $ R_{\text{s}} $ that achieves a pair of distortions $ (D_{\text{o}}, D_{\text{s}}) $ in the following lemma, whose proof is given in Ref. \cite{liu2021rate} and \cite{9844779}.
      
      \begin{Lemma} The minimum compression rate to achieve distortion pair $ (D_{\text{o}}, D_{\text{s}}) $,
       i.e.,  SRDF $ R(D_{\text{o}}, D_{\text{s}}) $, is given as
      	 \begin{equation}
  R(D_{\text{o}}, D_{\text{s}}) =  \min\limits_{ \substack{ P_{(\hat{X}, \hat{S})|X} \\ \textbf{E}\left[d_{\text{o}}(X, \hat{X})\right] \leq D_{\text{o}}  \\ \textbf{E}\left[\hat{d}_{\text{s}}(X, \hat{S})\right] \leq D_{\text{s}} }} H_{\text{KL}}\Big (P_{(X, \hat{X}, \hat{S})}|| P_X \times P_{(\hat{X}, \hat{S})} \Big) \label{def_R},
  \end{equation} 
  	 where design variable $ P_{(\hat{X}, \hat{S})|X} $  is the conditional distribution of $ (\hat{X}, \hat{S}) $ given $ X $ and $ P_X $ is the distribution of $ X $. Besides, $P_{(\hat{X}, \hat{S})}$ and $ P_{(X, \hat{X}, \hat{S})} $ are the joint distributions of $ (\hat{X}, \hat{S}) $ and $(X, \hat{X}, \hat{S})$, respectively, which can be computed by $P_X $ and $  P_{(\hat{X}, \hat{S})|X} $, and $H_{\text{KL}}(P_{(X, \hat{X}, \hat{S})}|| P_X \times P_{(\hat{X}, \hat{S})} ) $ is the Kulolback-Leibler (KL) distance between joint distribution $ P_{(X, \hat{X}, \hat{S})} $ and product distribution $ P_X \times P_{(\hat{X}, \hat{S})} $. Moreover,  $ \hat{d}_{\text{s}}(x, \hat{s}) = \sum_{s \in \mathcal{S} } P_{S|X}(s|x)d_{\text{s}}(s,\hat{s})$ is a distortion measure satisfying $ \hat{d}_{\text{s}}: \mathcal{X}\times\mathcal{\hat{S}}\to [0, + \infty) $, with $ P_{S|X} $ being the conditional distribution of $ S $ given $ X $, and its expected distorion $ \textbf{E}[\hat{d}_{\text{s}}(X, \hat{S})] $ is equivalent to $ \textbf{E}[d_{\text{s}}(S, \hat{S})] $.
      \end{Lemma}
      
       It is noted that the asymptotic optimality of semantic communications over the infinite discrete memoryless channel can always be obtained by SSCC\upcite{9844779,Thomas}. Then, under the considered SSCC framework shown in Fig. \ref{fig_system}, the distortion pair $ (D_{\text{o}}, D_{\text{s}}) $ is considered to be achievable if there exist separate source and channel codes such that the distortions between the transmitter and the receiver satisfy $ \textbf{E}[d_{\text{o}}(X, \hat{X})] \leq D_{\text{o}} $ and $ \textbf{E}[\hat{d}_{\text{s}}(X, \hat{S})] \leq D_{\text{s}}  $. Moreover,  we can immediately derive the following lemma to characterize the achievability of distortion pair $ (D_{\text{o}}, D_{\text{s}}) $.
       
       \begin{Lemma}
       \label{le_R_C}
       	 Utilizing the SSCC framework depicted in Fig. \ref{fig_system},  distortion pair $ (D_{\text{o}}, D_{\text{s}}) $ is achievable if and only if 
 \begin{equation}
 	R(D_{\text{o}}, D_{\text{s}}) \leq C,
 \end{equation}  	 
  	 where $ C $ is the channel capacity of the considered memoryless channel.
       \end{Lemma}


 \begin{Remark} \label{Re_property_R}
   Denote $ D_{\text{max}}^{\text{o}} $ as the maximum of observation distortion $D_{\text{o}} $, which is computed as the minimum expected distortion between $X$ and $ \hat{x} \in \hat{\mathcal{X}} $\upcite{yeung2008information},  i.e., $ D_{\text{max}}^{\text{o}}  = \min_{\hat{x}\in\hat{X}} \textbf{E}[d_{\text{o}}(X, \hat{x})] $. Similarly, define $ D_{\text{max}}^{\text{s}} $ as the maximum of semantic distortion $D_{\text{s}} $, i.e., $ D_{\text{max}}^{\text{s}}  = \min_{\hat{s}\in\hat{S}} \textbf{E}[\hat{d}_{\text{s}}(X, \hat{s})] $. Then, $ R(D_{\text{o}}, D_{\text{s}})$ has the following properties:
 	\begin{enumerate}
 		\item $ R(D_{\text{o}}, D_{\text{s}}) $ is jointly convex with $(D_{\text{o}}, D_{\text{s}}) $ and monotonically nonincreasing with respect to  $ D_{\text{o}}$ and $ D_{\text{s}} $. Moreover,  if $ D_{\text{o}} \geq D_{\text{max}}^{\text{o}} $ and $D_{\text{s}} \geq D_{\text{max}}^{\text{s}}$, it follows $ R(D_{\text{o}}, D_{\text{s}}) = 0 $.
 		\item  For any fixed $ D_{\text{s}} \in [0, D_{\text{max}}^{\text{s}}] $, there exists a $ D_{\text{o}}^\prime(D_{\text{s}}) \in [0, D_{\text{max}}^{\text{o}}]  $ such that $ R(D_{\text{o}}, D_{\text{s}}) = R_{\text{s}}(D_{\text{s}}) $  for all $ D_{\text{o}} \geq D_{\text{o}}^\prime(D_{\text{s}}) >0 $, where $ R_{\text{s}}(D_{\text{s}}) $ is the conventional rate-distortion function\upcite{cover1999elements} derived via the distortion between $ S $ and $ \hat{S} $. 
 		\item Similarly, for any fixed $ D_{\text{o}} \in [0, D_{\text{max}}^{\text{o}}] $, there exists a $ D_{\text{s}}^\prime(D_{\text{o}})  \in [0, D_{\text{max}}^{\text{s}}] $ such that $ R(D_{\text{o}}, D_{\text{s}}) = R_{\text{o}}(D_{\text{o}}) $  for all $ D_{\text{s}} \geq D_{\text{s}}^\prime(D_{\text{o}}) $, where $ R_{\text{o}}(D_{\text{o}}) $ is the conventional rate-distortion function derived via the distortion between $ X $ and $ \hat{X} $. 
 	\end{enumerate}
 \end{Remark} 	 
 \begin{IEEEproof}
    Property 1) can be easily proved by mimicking the proof for that of the conventional rate-distortion functions\upcite{cover1999elements}. The proofs of properties 2) and 3) can be found in  Appendix \ref{ap_Re_property_R}.
\end{IEEEproof}
\begin{Remark}
It is easy to observe that Remark \ref{Re_property_R} describes the trade-off among SRDF $R(D_{\text{o}}, D_{\text{s}})$ and distortions $ D_{\text{o}} $ and $ D_{\text{s}} $. More specifically, for any fixed semantic distortion $ D_{\text{s}} $, as the observation distortion $ D_{\text{o}} $ increases,  $ R(D_{\text{o}}, D_{\text{s}})$ first decreases, and then becomes a constant equal to the conventional rate-distortion function $ R_{\text{s}}(D_{\text{s}})$. This implies that there exists a lower bound of $ D_{\text{o}} $ for any fixed $ D_{\text{s}} $, denoted as $ D_{\text{o}}^\prime(D_{\text{s}}) $, such that  $ R(D_{\text{o}}, D_{\text{s}})$ is degenerated to $ R_{\text{s}}(D_{\text{s}})$ when $ D_{\text{o}} \geq D_{\text{o}}^\prime(D_{\text{s}}) $. Similarly, there also exists a lower bound $ D_{\text{s}}^\prime(D_{\text{o}}) $ of $ D_{\text{s}} $ such that  $ R(D_{\text{o}}, D_{\text{s}})$ is degenerated to the conventional rate-distortion function $ R_{\text{o}}(D_{\text{o}})$ when $ D_{\text{s}} \geq D_{\text{s}}^\prime(D_{\text{o}})$. Numerical results regarding SRDF $  R(D_{\text{o}}, D_{\text{s}}) $ are provided in Section \ref{num_result}, which show the monotonic and convex properties of $ R(D_{\text{o}}, D_{\text{s}})$ for some specific semantic sources.
\end{Remark}

However, there are still several challenges in effectively computing SRDF $ R(D_{\text{o}}, D_{\text{s}}) $ by \eqref{def_R}.
 	First, it is generally not possible to derive an expression for $ R(D_{\text{o}}, D_{\text{s}}) $  in closed form\upcite{9844779,yeung2008information}, and existing numerical methods for computing $ R(D_{\text{o}}, D_{\text{s}}) $  are limited to some specific sources, e.g., Gaussian\upcite{9844779} and binary sources\upcite{9834593}. Second, it is much more challenging to compute SRDF $R(D_{\text{o}}, D_{\text{s}})$ when semantic source distribution  $P_{(X,S)}$ is not perfectly known, i.e., only certain amount of samples of the semantic source pair $ (X,S) $ is available. To address these challenges, we propose two methods to compute $ R(D_{\text{o}}, D_{\text{s}}) $ based on the assumptions of different levels of semantic source distribution information.


   \section{Unknown Semantic Source Distributions}
   This section considers the case that the exact distribution information of semantic source $ (X,S) $ is not available, while its realizations are obtained from some large and high-dimensional datasets. First, we rewrite $ R(D_{\text{o}}, D_{\text{s}}) $ in \eqref{def_R} as an inf-sup form. Then,  we propose a neural-network-based method to compute $ R(D_{\text{o}}, D_{\text{s}}) $ and design a cascade neural network framework for a special case of  $ R(D_{\text{o}}, D_{\text{s}}) $ when $ S $ is a deterministic function of $X$.
   
   \subsection{Reformulation of $R(D_{\text{o}},D_{\text{s}})$}
   
    Before deriving the inf-sup form of $ R(D_{\text{o}}, D_{\text{s}}) $,  we define a semantic rate function, denoted as $R_1(Q_{(\hat{X}, \hat{S})}, D_{\text{o}}, D_{\text{s}})$, representing the compression rate, at which semantic source pair $ (X, S) $ is compressed to achieve distortion pair $(D_{\text{o}},D_{\text{s}})$, i.e.\upcite{dembo2002source,4567577,650987}, 
    \begin{equation}
    	R_1(Q_{(\hat{X}, \hat{S})}, D_{\text{o}}, D_{\text{s}}) \triangleq \min\limits_{ \substack{ P_{(\hat{X}, \hat{S})|X} \\ \textbf{E}\left[d_{\text{o}}(X, \hat{X})\right] \leq D_{\text{o}}  \\\textbf{E}\left[\hat{d}_{\text{s}}(X, \hat{S})\right] \leq D_{\text{s}}  }}  H_\text{KL}\Big(P_{(X, \hat{X}, \hat{S})}|| P_X \times Q_{(\hat{X}, \hat{S})}\Big). \label{R1}
    \end{equation}
    Here, $ Q_{(\hat{X}, \hat{S})} $ is a joint probability distribution  supported on the product alphabet  $ \mathcal{\hat{X}}\times \mathcal{\hat{S}} $.
    
    Next, we show that  $ R_1(Q_{(\hat{X}, \hat{S})}, D_{\text{o}}, D_{\text{s}}) $  has the following dual characterization.
    
             \begin{Proposition}
    \label{prop_R1}
    	The semantic rate function defined in \eqref{R1} can be equivalently computed as 
    	\begin{equation}
    		R_1(Q_{(\hat{X}, \hat{S})}, D_{\text{o}}, D_{\text{s}}) = \sup\limits_{\alpha_1, \alpha_2 \leq 0 } \alpha_1D_{\text{o}} + \alpha_2D_{\text{s}} - \Lambda_Q(\alpha_1, \alpha_2), \label{dual_R1}
     	\end{equation}
    \end{Proposition}
    with 
    \begin{equation}
    	\Lambda_Q(\alpha_1, \alpha_2) = \textbf{E}_{P_X}\Big[ \ln \textbf{E}_{Q_{(\hat{X}, \hat{S})}} e^{\alpha_1d_{\text{o}}(X, \hat{X})+ \alpha_2\hat{d}_{\text{s}}(X, \hat{S})  } \Big]. \label{Lambda_12}
    \end{equation}
        \begin{IEEEproof}
    	Please see Appendix \ref{ap_prop_R1}.
    \end{IEEEproof}

        \begin{Proposition} 
    \label{Prop_R_R1}
    	$ R(D_{\text{o}}, D_{\text{s}}) $ given in \eqref{def_R} can be equivalently computed as  
    	\begin{equation}
    		R(D_{\text{o}}, D_{\text{s}}) = \inf\limits_{Q_{(\hat{X}, \hat{S})}} R_1\big(Q_{(\hat{X}, \hat{S})}, D_{\text{o}}, D_{\text{s}}\big), \label{R_Q}
    	\end{equation}
    	and it has the same optimal solutions as problem \eqref{def_R}.
    \end{Proposition}
    \begin{IEEEproof}
    Please see Appendix \ref{ap_Prop_R_R1}.
    \end{IEEEproof}

 Together with Propositions \ref{prop_R1} and \ref{Prop_R_R1}, SRDF $ R(D_{\text{o}}, D_{\text{s}}) $ can be equivalently written as the following inf-sup problem
    \begin{equation}
    	R(D_{\text{o}}, D_{\text{s}}) = \inf\limits_{Q_{(\hat{X}, \hat{S})}}\sup\limits_{\alpha_1, \alpha_2 \leq 0 } \alpha_1D_{\text{o}} + \alpha_2D_{\text{s}} - \Lambda_Q(\alpha_1, \alpha_2). \label{exp_R}
    \end{equation}
 To solve this inf-sup problem, one straightforward idea is to first solve the inner supremum of \eqref{exp_R}, and then use the gradient descent method to optimize design variable $Q_{(\hat{X}, \hat{S})} $. As proved in Appendix \ref{ap_prop_R1}, objective function in \eqref{exp_R} is strictly concave with respect to $ \alpha_1 $ and $ \alpha_2 $, and thus possesses a unique solution to its inner supremum satisfying 
\begin{equation}
 	 D_{\text{o}} = \textbf{E}_{P_{X} \times Q_{(\hat{X}, \hat{S})}}\left[ d_{\text{o}}(X, \hat{X}) \frac{e^{\alpha_1d_{\text{o}}(X, \hat{X})+ \alpha_2\hat{d}_{\text{s}}(X, \hat{S})  }}{\textbf{E}_{Q_{(\hat{X}, \hat{S})}}\big[ e^{\alpha_1d_{\text{o}}(X, \hat{X})+ \alpha_2\hat{d}_{\text{s}}(X, \hat{S})  } \big] } \right],\label{1_part_1}
\end{equation} 
and
\begin{equation}
	D_{\text{s}} =\textbf{E}_{P_{X} \times Q_{(\hat{X}, \hat{S})}}\left[  \hat{d}_{\text{s}}(X, \hat{S})\frac{e^{\alpha_1d_{\text{o}}(X, \hat{X})+ \alpha_2\hat{d}_{\text{s}}(X, \hat{S})  }}{\textbf{E}_{Q_{(\hat{X}, \hat{S})}}\big[ e^{\alpha_1d_{\text{o}}(X, \hat{X})+ \alpha_2\hat{d}_{\text{s}}(X, \hat{S})  } \big] } \right],\label{1_part_2}
\end{equation}
where \eqref{1_part_1} and \eqref{1_part_2} are obtained by checking the first-order condition of the objective function in \eqref{exp_R}. However, it is difficult to get explicit expressions for $ \alpha_1 $ and $ \alpha_2 $ from \eqref{1_part_1} and \eqref{1_part_2}, making the computation of SRDF $ R(D_{\text{o}}, D_{\text{s}})$ challenging. Therefore, we present the following proposition to address this challenge.   

\begin{Proposition} 
\label{prop_Q}
	 For any fixed $ \alpha_1 $ and $ \alpha_2 $, $ \alpha_1, \alpha_2 \leq 0  $, considering 
	 \begin{equation}
	 	Q^{\star}_{(\hat{X}, \hat{S})} = \arg \inf\limits_{Q_{(\hat{X}, \hat{S})}} -\Lambda_Q(\alpha_1, \alpha_2), \label{pro_Lambda}
	 \end{equation}
	 SRDF $ R(D_{\text{o}}^{\star}, D_{\text{s}}^{\star}) $ given in \eqref{exp_R} is computed as
	 \begin{equation}
	 	R(D_{\text{o}}^{\star}, D_{\text{s}}^{\star}) = \alpha_1D_{\text{o}}^{\star} + \alpha_2D_{\text{s}}^{\star} - \Lambda_{Q^{\star}}(\alpha_1, \alpha_2), \label{R_star}
	 \end{equation}
	 and  $ D_{\text{o}}^{\star} $ and $ D_{\text{s}}^{\star} $ are obtained by replacing  $ Q_{(\hat{X}, \hat{S})} $ in \eqref{1_part_1} and \eqref{1_part_2} with $ Q^{\star}_{(\hat{X}, \hat{S})} $, respectively.

\end{Proposition}  
\begin{IEEEproof}
	Please see Appendix \ref{ap_prop_Q}.
\end{IEEEproof}  

\begin{Remark}
\label{remark_Q}
From Proposition \ref{prop_Q}, we observe that:
\begin{itemize}
	\item In the inf-sup problem \eqref{exp_R},  instead of focusing on directly solving $ \alpha_1 $ and $ \alpha_2 $,  we first fix $ \alpha_1 $ and $ \alpha_2  $, and then derive the optimal distribution $  Q^{\star}_{(\hat{X}, \hat{S})} $ by solving problem \eqref{pro_Lambda}. After that, with the obtained $  Q^{\star}_{(\hat{X}, \hat{S})} $,  $ D_{\text{o}}^{\star} $, $ D_{\text{s}}^{\star} $, and  $ R(D_{\text{o}}^{\star}, D_{\text{s}}^{\star}) $ can be computed by \eqref{1_part_1}, \eqref{1_part_2}, and \eqref{R_star}, accordingly. Finally, by selecting various values of $ \alpha_1 $ and $ \alpha_2 $, $ \alpha_1,\alpha_2 \leq 0 $, we then plot the whole $ R(D_{\text{o}}^{\star}, D_{\text{s}}^{\star}) $ surface.
	\item  However, similar to problem \eqref{def_R}, it is challenging to directly solve problem \eqref{pro_Lambda} by conventional optimization or statistical methods. First, design variable $ Q_{(\hat{X}, \hat{S})} $ has high-dimensional characteristics, which makes it difficult to be optimized. For example, for an 8-bit grayscale image dataset that consists of images with $m $ pixels in size and has $ n $ classes,  we consider that each realization of $ (\hat{X}, \hat{S}) $ is a reconstructed image and its label. Then, the full size of alphabet $ \mathcal{\hat{X}} $ is $|\mathcal{\hat{X}}|=2^{8m} $ and the dimension of $ Q_{(\hat{X}, \hat{S})} $ could be $ 2^{8m}n $. Second, it is difficult to directly compute $ \Lambda_Q(\alpha_1, \alpha_2) $ by \eqref{Lambda_12} since this section considers the case that semantic source distribution  $P_{(X,S)}$ is not perfectly known.  
	   \end{itemize}

\end{Remark}

\subsection{Neural-network-based Approach}

This subsection proposes a neural-network-based approach to solve problem \eqref{pro_Lambda}, and then derives the NESRD for $ R(D_{\text{o}}^{\star}, D_{\text{s}}^{\star}) $.  Specifically, to tackle the challenges discussed in Remark \ref{remark_Q}, we first design a generative neural network that takes a simple distribution, e.g., Gaussian distribution, as input, and aims to approximate distribution $Q_{(\hat{X}, \hat{S})}$ as closely as possible.  Then, parameters of the generative network are trained via gradient descent methods to minimize the objective function in \eqref{pro_Lambda}. Finally, SRDF $ R(D_{\text{o}}^{\star}, D_{\text{s}}^{\star}) $ given in \eqref{R_star} is estimated by samples of $(X,S)$ and the well-trained generative network. 
     
  First, we introduce a latent variable $ Z $ following distribution $ P_Z $ over alphabet $ \mathcal{Z} $ and then define a generative neural network from $ \mathcal{Z} $ to $ \mathcal{\hat{X}}\times \mathcal{\hat{S}} $ as $H(z, \boldsymbol{\theta})=[H^{(1)}(z, \boldsymbol{\theta}), H^{(2)}(z, \boldsymbol{\theta})]^T $, with $ H^{(1)}(z, \boldsymbol{\theta}) $ and $ H^{(2)}(z, \boldsymbol{\theta}) $ being realizations of reconstructed sources $ \hat{X} $ and $ \hat{S} $, respectively, and $ \boldsymbol{\theta} \in \Theta $ being the parameter to be optimized. After that,  we replace  $ Q_{(\hat{X}, \hat{S})} $ in \eqref{Lambda_12}  with  generative network $ H(Z, \boldsymbol{\theta}) $, which allows us to formulate problem \eqref{pro_Lambda} in terms of  parameter $\boldsymbol{\theta}$, i.e., problem \eqref{pro_Lambda} is transformed as 
 \begin{equation}
 	\inf\limits_{\boldsymbol{\theta} \in \Theta } \  -\textbf{E}_{P_X}\Big[ \ln \textbf{E}_{P_Z} e^{\alpha_1d_{\text{o}}\left(X, H^{(1)}(Z,\boldsymbol{\theta})\right)+ \alpha_2\hat{d}_{\text{s}}\left(X, H^{(2)}(Z,\boldsymbol{\theta})\right)  } \Big] . \label{H_R}
 \end{equation}

Then, we propose an iterative training algorithm to minimize problem \eqref{H_R} by optimizing parameter $ \boldsymbol{\theta} $ of the generative network $ H(Z, \boldsymbol{\theta}) $. We select i.i.d. samples $ x_1, \cdots, x_{N_1} $ from $ P_X $, i.i.d. samples $ s_{n_1}^{1}, \cdots, s_{n_1}^{N_2} $ from the conditional distribution $ P_{S|X=x_{n_1}} $ for each sample $ x_{n_1} $, $ n_1 = 1, \cdots, N_1$, and i.i.d. samples $ z_1, \cdots, z_M $ from distribution $ P_Z $, with $ N_1 $, $ N_2 $, and $ M $ being the number of observation samples, semantic state samples, and samples of latent variable $ Z $, respectively. Then, by approximating expectations in \eqref{H_R}  with  empirical averages over these samples of  $(X, S)$ and $ Z $, \eqref{H_R} is rewritten as
  \begin{align}
 	\inf\limits_{\boldsymbol{\theta} \in \Theta } \   &-\frac{1}{N_1} \sum_{n_1=1}^{N_1} \left [ \ln \frac{1}{M} \sum_{m=1}^M e^{\alpha_1d_{\text{o}}\left(x_{n_1}, H^{(1)}(z_m,\boldsymbol{\theta})\right) + \frac{\alpha_2}{N_2}\sum_{n_2 =1}^{N_2}d_{\text{s}}\left(s_{n_1}^{n_2}, H^{(2)}(z_m,\boldsymbol{\theta})\right)} \right ]. \label{L_theta}
 \end{align}
 Here, the objective function in \eqref{L_theta} is defined as the loss function $\mathcal{L}_{\boldsymbol{\theta}}(\alpha_1, \alpha_2) $  for training generative network $H(Z,\boldsymbol{\theta})$.  Besides, in order to ensure that the gradient of the loss function $ \nabla_{\boldsymbol{\theta}} \mathcal{L}_{\boldsymbol{\theta}} $ exists, distortion measures $ d_{\text{o}} $ and $ d_{\text{s}} $  must be differentiable. After that,  we proceed to update parameter $ \boldsymbol{\theta} $ by leveraging the gradient $ \nabla_{\boldsymbol{\theta}} \mathcal{L}_{\boldsymbol{\theta}} $ and employing the backpropagation algorithm\upcite{8054694}. 
 
By utilizing the well-trained generative network $ H(Z, \boldsymbol{\theta}^{\star}) $ with $ \boldsymbol{\theta}^{\star} $ being the trained parameters, we derive the NESRD for semantic source $ (X,S) $, which is summarized in the following proposition. 
\begin{Proposition}
	 For any fixed $ \alpha_1 $ and $ \alpha_2 $, $ \alpha_1, \alpha_2 \leq 0  $,  NESRD for semantic source $ (X,S) $ is given as \begin{align}
   \hat{R}_\Theta(\hat{D}_{\text{o}}^{\star}, \hat{D}_{\text{s}}^{\star}) \triangleq \alpha_1\hat{D}_{\text{o}}^{\star} + \alpha_2\hat{D}_{\text{s}}^{\star} + \mathcal{L}_{\boldsymbol{\theta}^{\star}}(\alpha_1, \alpha_2), \label{estimate_R} 
\end{align}
where $ \hat{D}_{\text{o}}^{\star}$ and $ \hat{D}_{\text{s}}^{\star} $ are computed as
\begin{align}
 	 	&\hat{D}_{\text{o}}^{\star} =\frac{1}{MN_1} \cdot \notag\\  &\sum_{m=1}^{M}\sum_{n_1=1}^{N_1} \left[ \frac{d_{\text{o}}\left(x_{n_1}, H^{(1)}(z_m,\boldsymbol{\theta}^{\star})\right)e^{\alpha_1d_{\text{o}}\left(x_{n_1}, H^{(1)}(z_m,\boldsymbol{\theta}^{\star})\right) + \frac{\alpha_2}{N_2}\sum_{n_2 =1}^{N_2}d_{\text{s}}\left(s_{n_1}^{n_2}, H^{(2)}(z_m,\boldsymbol{\theta}^{\star})\right)}}{\frac{1}{M} \sum_{m^{\prime}=1}^M e^{\alpha_1d_{\text{o}}\left(x_{n_1}, H^{(1)}(z_m^{\prime},\boldsymbol{\theta}^{\star})\right) + \frac{\alpha_2}{N_2}\sum_{n_2 =1}^{N_2}d_{\text{s}}\left(s_{n_1}^{n_2}, H^{(2)}(z_m^{\prime},\boldsymbol{\theta}^{\star})\right)} } \right], \label{average_part_1} 
\end{align} 
and
\begin{align}
	&\hat{D}_{\text{s}}^{\star} =\frac{1}{MN_1N_2} \cdot \notag\\  &\sum_{m=1}^{M}\sum_{n_1=1}^{N_1}\sum_{n_2 =1}^{N_2} \left[ \frac{ d_{\text{s}}\left(s_{n_1}^{n_2}, H^{(2)}(z_m,\boldsymbol{\theta}^{\star})\right)e^{\alpha_1d_{\text{o}}\left(x_{n_1}, H^{(1)}(z_m,\boldsymbol{\theta}^{\star})\right) + \frac{\alpha_2}{N_2}\sum_{n_2^{\prime} =1}^{N_2}d_{\text{s}}\left(s_{n_1}^{n_2^{\prime}}, H^{(2)}(z_m,\boldsymbol{\theta}^{\star})\right)}}{\frac{1}{M} \sum_{m^{\prime}=1}^M e^{\alpha_1d_{\text{o}}\left(x_{n_1}, H^{(1)}(z_m^{\prime},\boldsymbol{\theta}^{\star})\right) + \frac{\alpha_2}{N_2}\sum_{n_2^{\prime} =1}^{N_2}d_{\text{s}}\left(s_{n_1}^{n_2^{\prime}}, H^{(2)}(z_m^{\prime},\boldsymbol{\theta}^{\star})\right)} } \right],\label{average_part_2}
\end{align}
respectively.
\end{Proposition} 
\begin{IEEEproof}
	It is easy to see that \eqref{estimate_R}-\eqref{average_part_2} are obtained by  replacing  $ Q_{(\hat{X}, \hat{S})} $ in \eqref{1_part_1} and \eqref{1_part_2} with $ H(Z, \boldsymbol{\theta}^{\star}) $,   approximating expectations in \eqref{1_part_1} and \eqref{1_part_2}  with  empirical averages over training samples given in \eqref{L_theta},  and replacing $ \Lambda_{Q^{\star}}(\alpha_1, \alpha_2) $ in \eqref{R_star} with $ -\mathcal{L}_{\boldsymbol{\theta}^{\star}}(\alpha_1, \alpha_2) $ defined in \eqref{L_theta}.
\end{IEEEproof}

In conclusion, we summarize the neural-network-based approach for NESRD in Algorithm \ref{alg_NESRD}.

   \begin{algorithm}[htbp]

	\caption{Neural-network-based approach for NESRD. \label{alg_NESRD}}
	\label{Joint Parameter Estimation And Data Detection} 
	\begin{algorithmic}[1]
	\Require  Parameters $ \alpha_1  $ and  $ \alpha_2 $ satisfying $ \alpha_1, \alpha_2 \leq 0 $, and the number $T$ of training steps.
	\Ensure $ (\hat{D}_{\text{o}}^{\star}, \hat{D}_{\text{s}}^{\star}, \hat{R}_\Theta(\hat{D}_{\text{o}}^{\star}, \hat{D}_{\text{s}}^{\star})) $.
    \State Set a generative neural network $ H(z, \boldsymbol{\theta})=[H^{(1)}(z, \boldsymbol{\theta}), H^{(2)}(z, \boldsymbol{\theta})]^T:\mathcal{Z} \to \mathcal{\hat{X}}\times \mathcal{\hat{S}}  $;
    \State \textbf{For} {$ t= 1,2,\cdots, T $} \textbf{do} 

    \State \quad Choose batch of $ \hat{N}_1 $ i.i.d. observation samples $ \{ x_1, \cdots, x_{\hat{N}_1} \}$ from $ P_X $;
    \State \quad Choose batch of $ \hat{N}_2 $ i.i.d. semantic state samples $ \{s_{\hat{n}_1}^{1}, \cdots, s_{\hat{n}_1}^{\hat{N}_2}\} $ from the conditional distribution $ P_{S|X=x_{\hat{n}_1}} $ for each sample $ x_{\hat{n}_1} $ with $ \hat{n}_1 = 1, \cdots, \hat{N}_1$;
    \State \quad Choose batch of $ \hat{M} $ i.i.d. samples $ \{ z_1, \cdots, z_{\hat{M}} \}$ that are generated from $ P_Z $;
    \State \quad Calculate loss function $  \mathcal{L}_{\boldsymbol{\theta}}(\alpha_1, \alpha_2) $ in \eqref{L_theta}; 
    \State \quad Utilize gradient $ \nabla_{\boldsymbol{\theta}} \mathcal{L}_{\boldsymbol{\theta}}(\alpha_1, \alpha_2) $ for backpropagation\upcite{8054694};

    \State \textbf{End for}
    \State Calculate  $ \hat{D}_{\text{o}}^{\star}$, $\hat{D}_{\text{s}}^{\star}$, and $\hat{R}_\Theta(\hat{D}_{\text{o}}^{\star}, \hat{D}_{\text{s}}^{\star}) $ by \eqref{estimate_R}-\eqref{average_part_2}. 
	\end{algorithmic}
\end{algorithm}

	Moreover, the following corollary shows that NESRD $ \hat{R}_\Theta(\hat{D}_{\text{o}}^{\star}, \hat{D}_{\text{s}}^{\star})$ converges almost surely to  SRDF $R(D_{\text{o}}^{\star}, D_{\text{s}}^{\star})$ given in \eqref{R_star} as the number of samples goes to infinity, i.e.,  $ \hat{R}_\Theta(\hat{D}_{\text{o}}^{\star}, \hat{D}_{\text{s}}^{\star})$ is a strongly consistent estimator for $ R(D_{\text{o}}^{\star}, D_{\text{s}}^{\star}) $. Notably, the proof of this corollary is actually the same as that for the conventional rate-distortion function\upcite{10124059}.
\begin{Corollary}
	(Strong consistency of NESRD). Considering that the distribution $ P_Z $ of the latent variable $Z $ is continuous with respect to Lebesgue measure, and the distortion measures $ \hat{d}_{\text{s}} $ and $ d_{\text{o}} $ are both $ L_d $-Lipschitz,  $ \hat{R}_\Theta(\hat{D}_{\text{o}}^{\star}, \hat{D}_{\text{s}}^{\star})$ in \eqref{estimate_R} is a strongly consistent estimator of $ R(D_{\text{o}}^{\star}, D_{\text{s}}^{\star}) $ given in \eqref{R_star}, i.e., 
	\begin{equation}
	    \text{Pr}\left\{\lim\limits_{ N_1, N_2,  M\to\infty} \left(\hat{D}_{\text{o}}^{\star}, \hat{D}_{\text{s}}^{\star}, \hat{R}_\Theta(\hat{D}_{\text{o}}^{\star}, \hat{D}_{\text{s}}^{\star}) \right) = \left(D_{\text{o}}^{\star}, D_{\text{s}}^{\star}, R(D_{\text{o}}^{\star}, D_{\text{s}}^{\star})\right)\right\} = 1. 
	\end{equation}
\end{Corollary}

\begin{Remark}
	Although NESRD has been shown in Corollary 3.1 to be a strongly consistent estimator of
SRDF, it is essential to note that its estimation inaccuracy is sensitivity to the number of samples. More specifically, it has been proved in Ref.\cite{10124059} and \cite{pmlr-v108-mcallester20a} that NESRD $ \hat{R}_\Theta(\hat{D}_{\text{o}}^{\star}, \hat{D}_{\text{s}}^{\star})$ cannot be larger than $O(\log N_1N_2)$ with $ N_1N_2 $ being the number of semantic source samples in \eqref{estimate_R}.
\end{Remark}

\begin{Remark}
	 It is easy to see that Algorithm \ref{alg_NESRD} can be utilized to estimate SRDFs for the case with perfectly known semantic source distributions. In such cases, the required i.i.d. semantic source samples in \eqref{L_theta} can be directly generated from the known source distribution $ P_{(X,S)} $. Then, the corresponding NESRD can be obtained by Algorithm \ref{alg_NESRD} accordingly.   
\end{Remark}

\subsection{A Special Case for NESRD}
    \begin{figure}[htbp]
	\centering
	\includegraphics[width=3.5in]{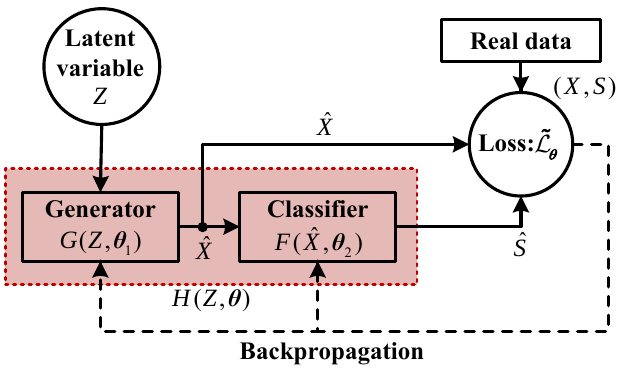}
	\caption{Training diagram for NESRD with $ S = h(X) $.}
	\label{nn_model}
    \end{figure}
 In this subsection, we mainly analyze a special case for SRDF $ R(D_{\text{o}}, D_{\text{s}}) $, where the intrinsic semantic state $ S $ is a deterministic function of the extrinsic observation $ X $, i.e., the semantic source $ (X,S)$ is obtained from labeled datasets, e.g., MNIST, SVHN, and CIFAR-10 datasets. In such datasets,  the raw images are samples of extrinsic observation $X$, while their corresponding labels can be regarded as samples of intrinsic semantic state $S$. For the ease of analysis, it follows $ S = h(X) $, and accordingly, semantic distortion measure $ \hat{d}_{\text{s}}(x, \hat{s}) $ in \eqref{def_R} can be simplified as
 \begin{equation}
 	\hat{d}_{\text{s}}(x, \hat{s}) = \sum_{s \in \mathcal{S} } p_{S|X}(s|x)d_{\text{s}}(s,\hat{s}) = d_{\text{s}}(h(x), \hat{s}). \label{d_f_x}
 \end{equation} 
 
 Moreover, we further design a cascade neural network framework to estimate SRDF for the case with $ S = h(X) $, which is depicted in Fig. \ref{nn_model}. Under this framework, generative network $ H(Z, \boldsymbol{\theta}) $ is composed of a generator $ G(Z,\boldsymbol{\theta}_1)$ and a classifier $ F(\hat{X}, \boldsymbol{\theta}_2) $, $\boldsymbol{\theta}=[\boldsymbol{\theta}_1, \boldsymbol{\theta}_2]^T$, with the generator producing samples of reconstructed observation $ \hat{X} $, and the classifier generating samples of reconstructed semantic state $ \hat{S} $. Besides, since $ S $ is a function of $ X $, we consider the case that $ \hat{S} $ can also be regarded as a function of $ \hat{X} $ so that its distribution $ P_{\hat{S}} $ can be directly estimated by  $ \hat{X} $. Therefore, the generator is connected with the classifier in series, and the generative network $H(Z, \boldsymbol{\theta})$ defined in \eqref{H_R} is now equivalent to 
\begin{equation}
	H(Z, \boldsymbol{\theta})  = \left[G(Z,\boldsymbol{\theta}_1), F\left(G(Z,\boldsymbol{\theta}_1), \boldsymbol{\theta}_2\right)\right]^T. \label{h_f_x}
\end{equation}
Moreover, together with \eqref{d_f_x} and \eqref{h_f_x}, the loss function in \eqref{L_theta} for training parameters $ \boldsymbol{\theta} $ can be simplified as
\begin{equation}
\tilde{\mathcal{L}}_{\boldsymbol{\theta}}(\alpha_1, \alpha_2) = -\frac{1}{N_1} \sum_{n_1=1}^{N_1} \left [ \ln \frac{1}{M} \sum_{m=1}^{M} e^{\alpha_1d_{\text{o}}\left(x_{n_1}, G(z_m,\boldsymbol{\theta}_1)\right) + \alpha_2 d_{\text{s}}\left(s_{n_1}, F(G(z_m, \boldsymbol{\theta}_1),\boldsymbol{\theta}_2)\right)} \right ].  \label{L_tilde}
\end{equation}
Here,  $ (x_1, s_1), \cdots, (x_{N_1}, s_{N_1}) $ are i.i.d. sample pairs from the labeled dataset. First, generator $ G(Z,\boldsymbol{\theta}_1)$ and classifier $ F(\hat{X}, \boldsymbol{\theta}_2) $ are both pre-trained by the labeled dataset.  Then, similar to the training procedure in Algorithm \ref{alg_NESRD}, we iteratively update parameters $ \boldsymbol{\theta} $ by utilizing the gradient of the loss function $ \nabla_{\boldsymbol{\theta}} \tilde{\mathcal{L}}_{\boldsymbol{\theta}}(\alpha_1, \alpha_2) $ and the backpropagation method\upcite{8054694}. Finally, as the training algorithm converges,   $ \hat{D}_{\text{o}}^{\star}$, $\hat{D}_{\text{s}}^{\star}$, and $\hat{R}_\Theta(\hat{D}_{\text{o}}^{\star}, \hat{D}_{\text{s}}^{\star}) $ are calculated by \eqref{estimate_R}-\eqref{average_part_2}, with $\hat{d}_{\text{s}}(\cdot)$, $ H(\cdot) $, and $ \mathcal{L}_{\boldsymbol{\theta}}(\cdot) $ in \eqref{estimate_R}-\eqref{average_part_2} being replaced with \eqref{d_f_x}, \eqref{h_f_x}, and \eqref{L_tilde}, respectively.
In conclusion, we summarize the training algorithm for estimating the SRDF for labeled datasets in Algorithm \ref{alg_s_NESRD}.

   \begin{algorithm}[htbp]
	\caption{Training algorithm for NESRD with the labeled dataset.\label{alg_s_NESRD}}
	\label{Joint Parameter Estimation And Data Detection} 
	\begin{algorithmic}[1]
	\Require  Parameters $ \alpha_1  $ and  $ \alpha_2 $ satisfying $ \alpha_1, \alpha_2 \leq 0 $, the labeled dataset $ \mathcal{K} $, and the number $T$ of training steps.
	\Ensure $ (\hat{D}_{\text{o}}^{\star}, \hat{D}_{\text{s}}^{\star}, \hat{R}_\Theta(\hat{D}_{\text{o}}^{\star}, \hat{D}_{\text{s}}^{\star})) $.
	\State Utilize dataset $\mathcal{K}$ to pre-train generator $ G $ in the same way as training the generative network in GAN\upcite{Goodfellow,8253599,8039016};
	\State Pre-train classifier $F$ with dataset $ \mathcal{K} $; 
	\State \textbf{For} {$ t= 1,2,\cdots, T $} \textbf{do} 
    \State \quad Choose batch of $ \hat{N}_1 $ semantic source sample couples  $ \{ (x_1, s_1), \cdots, (x_{\hat{N}_1}, s_{\hat{N}_1}) \}$ from dataset $ \mathcal{K} $;

    \State \quad Choose batch of $ \hat{M} $ samples $ \{ z_1, \cdots, z_{\hat{M}} \}$ that are generated from $ P_Z $;
    \State \quad Calculate loss function $ \tilde{\mathcal{L}}_{\boldsymbol{\theta}}(\alpha_1, \alpha_2) $ by \eqref{L_tilde}; 
    \State \quad Utilize gradient $ \nabla_{\boldsymbol{\theta}}\tilde{\mathcal{L}}_{\boldsymbol{\theta}}(\alpha_1, \alpha_2) $ for backpropagation;
    \State \textbf{End for}
    \State Utilizing the trained generator $ G $ and classfier $F $,  calculate $ \hat{D}_{\text{o}}^{\star}$, $\hat{D}_{\text{s}}^{\star}$, and $\hat{R}_\Theta(\hat{D}_{\text{o}}^{\star}, \hat{D}_{\text{s}}^{\star}) $ by \eqref{estimate_R}-\eqref{average_part_2}, where $\hat{d}_{\text{s}}(\cdot)$, $ H(\cdot) $, and $ \mathcal{L}_{\boldsymbol{\theta}}(\cdot) $ in \eqref{estimate_R}-\eqref{average_part_2} are substituted with \eqref{d_f_x}, \eqref{h_f_x}, and \eqref{L_tilde}, respectively.
	\end{algorithmic}
\end{algorithm}  

 \section{Perfectly Known Semantic Source Distributions }   \label{sec_ba}
        This section considers the case that the distribution of semantic source pair $(X,S)$ is completely known and has a discrete form\footnote{When semantic source distribution $ P_{(X,S)} $ is not discrete, we can approximate it with a discrete distribution by discretizing the possible values of $(X,S)$ into a finite number of points and estimating their corresponding probability mass functions via the known distribution $ P_{(X,S)} $.}, and we generalize the conventional BA algorithm\upcite{Arimoto,Blahut} to numerically compute SDRF  $ R(D_{\text{o}}, D_{\text{s}}) $. Besides, we analyze the computational complexity of the proposed general BA algorithm.
\subsection{General BA Algorithm}

 
  To derive the general BA algorithm, we first show that computing $ R(D_{\text{o}}, D_{\text{s}}) $ in \eqref{def_R}  is equivalent to solving the following double minimum problem.   
  \begin{Lemma}
  For any  fixed $ \lambda_1 \leq 0 $ and  $ \lambda_2 \leq 0 $, there exist $ \hat{D}_{\text{o}} $ and $ \hat{D}_{\text{s}} $ satisfying 
\begin{equation}
  \lambda_1 = \frac{\partial R(D_{\text{o}}, \hat{D}_{\text{s}})}{\partial D_{\text{o}}}\bigg|_{D_{\text{o}} = \hat{D}_{\text{o}}}	\text{and},  \quad  \lambda_2 = \frac{\partial R(\hat{D}_{\text{o}},D_{\text{s}})}{\partial D_{\text{s}}}\bigg|_{D_{\text{s}} = \hat{D}_{\text{s}}} \label{alpha_condition}
  \end{equation}  
   such that
  	\begin{align}
  		R(\hat{D}_{\text{o}},\hat{D}_{\text{s}}) - \lambda_1\hat{D}_{\text{o}} - \lambda_2\hat{D}_{\text{s}}  = & \min\limits_{P_{(\hat{X}, \hat{S})|X}>0}\min\limits_{\hat{P}_{(\hat{X}, \hat{S})}>0}\bigg[ \sum\limits_{x,\hat{x},\hat{s}}P_{(X, \hat{X}, \hat{S})}(x, \hat{x},\hat{s})\ln\frac{P_{(\hat{X}, \hat{S})|X}(\hat{x},\hat{s}|x)}{\hat{P}_{(\hat{X}, \hat{S})}(\hat{x}, \hat{s})} \notag \\  & - \sum\limits_{x,\hat{x},\hat{s}}P_X(x)P_{(\hat{X}, \hat{S})|X}(\hat{x},\hat{s}|x)(\lambda_1d_{\text{o}}(x,\hat{x})+\lambda_2\hat{d}_{\text{s}}(x,\hat{s}))\bigg], 
  		\label{double_inf_R}
  	\end{align}
  	where $ \hat{P}_{(\hat{X}, \hat{S})} $ is any distribution supported over $   \mathcal{\hat{X}}\times \mathcal{\hat{S}}$.
  \end{Lemma}
  \begin{IEEEproof}
  	The proof to this lemma is the same as that for the conventional rate-distortion functions; see, e.g., Ref.\cite{yeung2008information} and \cite{cover1999elements}. A similar result was obtained in Ref. \cite{9834593}.
  \end{IEEEproof}
 
   Then, by mimicking the conventional BA method given in Ref.\cite{yeung2008information},   we obtain the general BA algorithm by iteratively solving the double minimum in \eqref{double_inf_R}, which is summarized in Algorithm \ref{alg_ba}. In this algorithm, we first arbitrarily choose a strictly positive transition distribution  $ P_{(\hat{X}, \hat{Sx})|X}^{(0)} $ as an initial point. Then,  joint distribution $ \hat{P}_{(\hat{X}, \hat{S})}^{(k)} $ and conditional distribution $ P_{(\hat{X}, \hat{S})|X}^{(k+1)} $ are recursively computed as \begin{equation}
   	\hat{P}_{(\hat{X}, \hat{S})}^{(k)}(\hat{x},\hat{s}) = \sum\limits_{x}P_X(x) P_{(\hat{X}, \hat{S})|X}^{(k)}(\hat{x},\hat{s}|x),
   	\label{eq_P}
   \end{equation} 
   \begin{equation}
    P_{(\hat{X}, \hat{S})|X}^{(k+1)}(\hat{x},\hat{s}|x) = \frac{\hat{P}_{(\hat{X}, \hat{S})}^{(k)}(\hat{x},\hat{s})e^{\lambda_1d_{\text{o}}(x,\hat{x})+\lambda_2\hat{d}_{\text{s}}(x,\hat{s})}}{\sum\limits_{\hat{x}^\prime,\hat{s}^\prime}	\hat{P}_{(\hat{X}, \hat{S})}^{(k)}(\hat{x}^\prime,\hat{s}^\prime)e^{\lambda_1d_{\text{o}}(x,\hat{x}^\prime)+\lambda_2\hat{d}_{\text{s}}(x,\hat{s}^\prime)}},
    \label{eq_c_P}
   \end{equation} 
 respectively, where \eqref{eq_P} and \eqref{eq_c_P} are the generalizations of those iteration steps in the conventional BA method\upcite{yeung2008information}. Moreover,
  as the iteration index $k$ goes to infinity, we have 
   \begin{equation}
   	\left(\textbf{E}_{P_{(X, \hat{X})}^{(k)}}[d_{\text{o}}(X, \hat{X})], \textbf{E}_{P_{(X, \hat{S})}^{(k)}}[\hat{d}_{\text{s}}(X, \hat{S})],  H_{\text{KL}}\left(P_{(X, \hat{X}, \hat{S})}^{(k)}|| P_X \times P_{(\hat{X}, \hat{S})}^{(k)} \right) \right) \to \left(\hat{D}_{\text{o}},\hat{D}_{\text{s}}, R(\hat{D}_{\text{o}}, \hat{D}_{\text{s}})\right),
   	\label{converge_R}
   \end{equation}
 where $ P_{(X, \hat{X},  \hat{S})}^{(k)} $ is calculated as $ P_{(X, \hat{X},  \hat{S})}^{(k)}(x,\hat{x},\hat{s}) = P_X(x) P_{(\hat{X}, \hat{S})|X}^{(k)}(\hat{x},\hat{s}|x) $ for each triple $ (x,\hat{x},\hat{s})\in \mathcal{X}\times \mathcal{\hat{X}}\times \mathcal{\hat{S}} $, and  $ P_{(X, \hat{X})}^{(k)} $ and $ P_{(X, \hat{S})} $ are computed as $ P_{(X, \hat{X})}^{(k)}(x,\hat{x}) = \sum\limits_{\hat{s}} P_{(X, \hat{X},  \hat{S})}^{(k)}(x,\hat{x},\hat{s})$ and $ P_{(X, \hat{S})}(x,\hat{s})=  \sum\limits_{\hat{x}} P_{(X, \hat{X},  \hat{S})}^{(k)}(x,\hat{x},\hat{s})$, respectively. Besides, the convergence in \eqref{converge_R} is obtained similar to that of the conventional BA algorithm\upcite{yeung2008information}.
 
    \begin{algorithm}[htbp]
	\caption{General BA algorithm for computing SDRF  $ R(\hat{D}_{\text{o}},\hat{D}_{\text{s}}) $. \label{alg_ba}}
	\begin{algorithmic}[1]
	\Require  $ P_{(X,S)} $.
	\Ensure $ (\hat{D}_{\text{o}},\hat{D}_{\text{s}}, R(\hat{D}_{\text{o}},\hat{D}_{\text{s}}))$.
    \State Set $ \lambda_1 $, $ \lambda_2 <=0 $; 
    \State Arbitrarily choose a strictly positive transition distribution  $ P_{(\hat{X}, \hat{S})|X}^{(0)} >0 $; 
    \State Set $ k = 0 $;
    \State \textbf{While} {$k\leq K $} \textbf{do}
    \State \quad Compute $ \hat{P}_{(\hat{X}, \hat{S})}^{(k)} $ and $ P_{(\hat{X}, \hat{S})|X}^{(k+1)} $  by  \eqref{eq_P} and \eqref{eq_c_P}, respectively;
    \State \quad $k = k+1$;
    \State \textbf{End while}
     \State Compute  $ \hat{D}_{\text{o}} $, $ \hat{D}_{\text{s}} $, and $R(\hat{D}_{\text{o}}, \hat{D}_{\text{s}})$ by  $\hat{D}_{\text{o}}=\textbf{E}_{P_{(X,\hat{X})}^{(K)}}[d_{\text{o}}(X, \hat{X})]$, $ \hat{D}_{\text{s}}=\textbf{E}_{P_{(X,  \hat{S})}^{(K)}}[\hat{d}_{\text{s}}(X, \hat{S})]$, and $ R(\hat{D}_{\text{o}}, \hat{D}_{\text{s}})=H_{\text{KL}}\left(P_{(X, \hat{X}, \hat{S})}^{(K)}|| P_X \times \hat{P}_{(\hat{X}, \hat{S})}^{(K)}\right) $, respectively.
	\end{algorithmic}
\end{algorithm}   
\subsection{Computational Complexity }
 In this subsection, we analyze the computational complexity of the proposed general BA algorithm.
 \begin{enumerate}
 	\item To compute $ \hat{P}_{(\hat{X}, \hat{S})}^{(k)}(\hat{x},\hat{s}) $ in \eqref{eq_P}, we need $ \left|\mathcal{X}\right| $ multiplications and $ |\mathcal{X} |-1 $ additions for a specific tuple $(\hat{x},\hat{s})  $, thus deriving $ O(\left|\mathcal{X}\right|) $ operations. As a result, we need $ O(|\mathcal{X}||\mathcal{\hat{X}}||\mathcal{\hat{S}}|) $ operations for all $ (\hat{x},\hat{s}) $ tuples.
 	\item To compute  $  P_{(\hat{X}, \hat{S})|X}^{(k+1)}(\hat{x},\hat{s}|x) $ shown in \eqref{eq_c_P}, we consider the computations for exponentiation, $ d_{\text{s}}(s,\hat{s}) $, and $ d_{\text{o}}(x, \hat{x}) $, and all these operations need constant numbers of operations, i.e., $ O(1) $.  Then, $ \hat{d}_{\text{s}}(x,\hat{s}) $ is a summation over $ | \mathcal{S} | $ variables, which requires $ O(| \mathcal{S} | ) $ computations for a specific tuple $ (x, \hat{s}) $. Moreover, the numerator in \eqref{eq_c_P} needs  $ O(| \mathcal{S} | ) $ computations, and  the denominator in \eqref{eq_c_P} requires $O(|\mathcal{S}||\mathcal{\hat{X}}||\mathcal{\hat{S}}|)  $ since it is a summation over all $ (\hat{x},\hat{s}) $ tuples. Therefore, computing $  P_{(\hat{X}, \hat{S})|X}^{(k+1)}(\hat{x},\hat{s}|x) $ needs $O(|\mathcal{X}||\mathcal{S}||\mathcal{\hat{X}}|^2|\mathcal{\hat{S}}|^2)  $ computations for all $ (x,\hat{x},\hat{s}) $ tuples.
 \end{enumerate}
 In conclusion, the computational complexity of the proposed BA algorithm is $O(|\mathcal{X}||\mathcal{S}||\mathcal{\hat{X}}|^2|\mathcal{\hat{S}}|^2)  $ for each iteration.
 
It is easy to see that $|\mathcal{X}|$ grows exponentially with respect to the dimension of $X$, leading to an exponential growth in the computational complexity of the proposed general BA algorithm. This drawback also applies to $S$. Moreover, semantic source pair $(X, S)$ is typically derived from large datasets in practical applications and exhibits high-dimensional characteristics, rendering the general BA algorithm inefficient in such cases.

\section{Numercial and Simulation Results} \label{num_result}
In this section, we present some experimental and simulation results to validate the proposed NESRD for various typical semantic sources. First, we examine the case when $ X $ and $ S $ are jointly Gaussian, and compare the proposed neural-network-based approach with the proposed general BA algorithm and the semi-definite programming (SDP) method\upcite{9844779}. Then, we calculate the NESRD for some classical image datasets, e.g., MNIST and SVHN datasets, and compare it with the hyperprior-based compression method\upcite{balle2018variational}, which adopted as the benchmark for the DNN-based image compression schemes.

\subsection{Joint Gaussian Source}
\subsubsection{Datasets}
 In this subsection, extrinsic observation $ X $ follows a multivariate Gaussian distribution $ \mathcal{N}(0, K_X) $, and intrinsic semantic state $ S $ satisfies $ S = HX + W $, where $ H $ is a constant matrix and $ W $ is also a Gaussian random vector following $\mathcal{N}(0, K_W) $. $ K_X $, $ H $, and $ K_W $ are set as\upcite{liu2021rate}
\begin{equation}
K_X = \left[
	\begin{array}{cccc}
	 11 & 0 & 0.5 \\
	 0 & 3 & -2 \\
	 0.5 & -2 & 2.35
	\end{array}
	\right], H = \left[
	\begin{array}{cccc}
		0.0701& 0.305 & 0.457 \\
		-0.0305 & -0.220 & 0.671
	\end{array}
	\right], K_W=\left[
	\begin{array}{cccc}
		0.71 & -0.305 \\
		-0.305& 0.220
	\end{array}
	\right]. \label{para_gaussian}
\end{equation}
To leverage Algorithm \ref{alg_NESRD} for NESRD of the joint Gaussian source $(X,S)$, we generate a sample set as follows: First, we independently obtain  $ N_1 =50000 $ samples of $ X $ following the distribution $  \mathcal{N}(0, K_X)$. Then, for each sample $ x_{n_1} $, $ n_1 = 1,\cdots, 50000$, we select $ N_2 = 1000 $ samples of $ S $ which are drawn i.i.d. from the conditional distribution $ P(S|X=x_{n_1}) \sim \mathcal{N}(Hx_{n_1}, K_W) $.

\subsubsection{Training settings}
In Algorithm \ref{alg_NESRD}, the distribution of the latent variable $ Z $ is set to be $ P_Z = \mathcal{N}(0, I_{10}) $ with $ 10 $ being the dimension of the latent space, and the generative network $ H(z, \boldsymbol{\theta})=[H^{(1)}(z, \boldsymbol{\theta}), H^{(2)}(z, \boldsymbol{\theta})]^T:\mathcal{Z} \to \mathcal{\hat{X}}\times \mathcal{\hat{S}} $ is parameterized by a $3$-layer fully connected neural network with  $5$  units in both the hidden and output layers: the first two dimensions of the output are used to represent $ \hat{S} $, and the last three dimensions are used to represent $ \hat{X} $. Besides, the learning rate of the considered neural network is set as $ 1\times10^{-4} $, and the training epoch is set as $50$. Moreover, we utilize the squared-error distortion measures for both the intrinsic semantic state and the extrinsic observation, i.e., $ d_{\text{s}}(s,\hat{s})= \Vert s- \hat{s}\Vert_2^2  $ and $ d_{\text{o}}(x,\hat{x})= \Vert x- \hat{x}\Vert_2^2 $.  Based on the above parameters, we apply Algorithm \ref{alg_NESRD} to estimate SRDF $ R(D_{\text{o}}, D_{\text{s}}) $ for the considered joint Gaussian source.
\subsubsection{Experiments}

\begin{figure}[htbp] 
\centering
\subfigure[ Surface of  $ \hat{R}_\Theta(D_{\text{o}}, D_{\text{s}}) $ ]{
\includegraphics[width=2.85in]{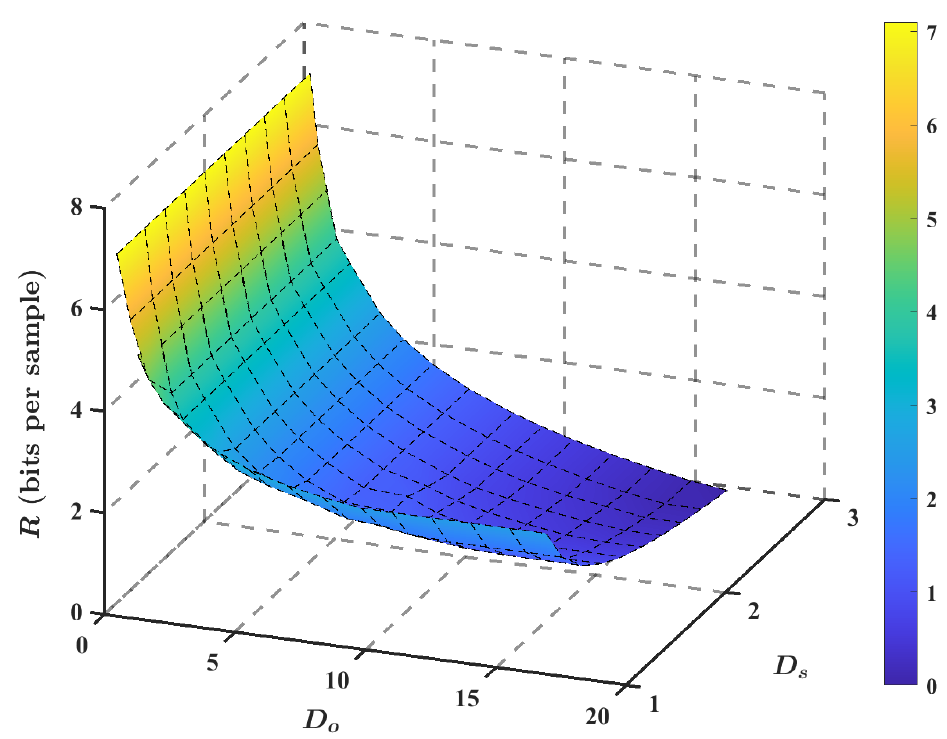}}
 \hspace{0.51in}
\subfigure[ \label{fig_contour_G} Contour of  $ \hat{R}_\Theta(D_{\text{o}}, D_{\text{s}}) $]{
\includegraphics[width=2.85in]{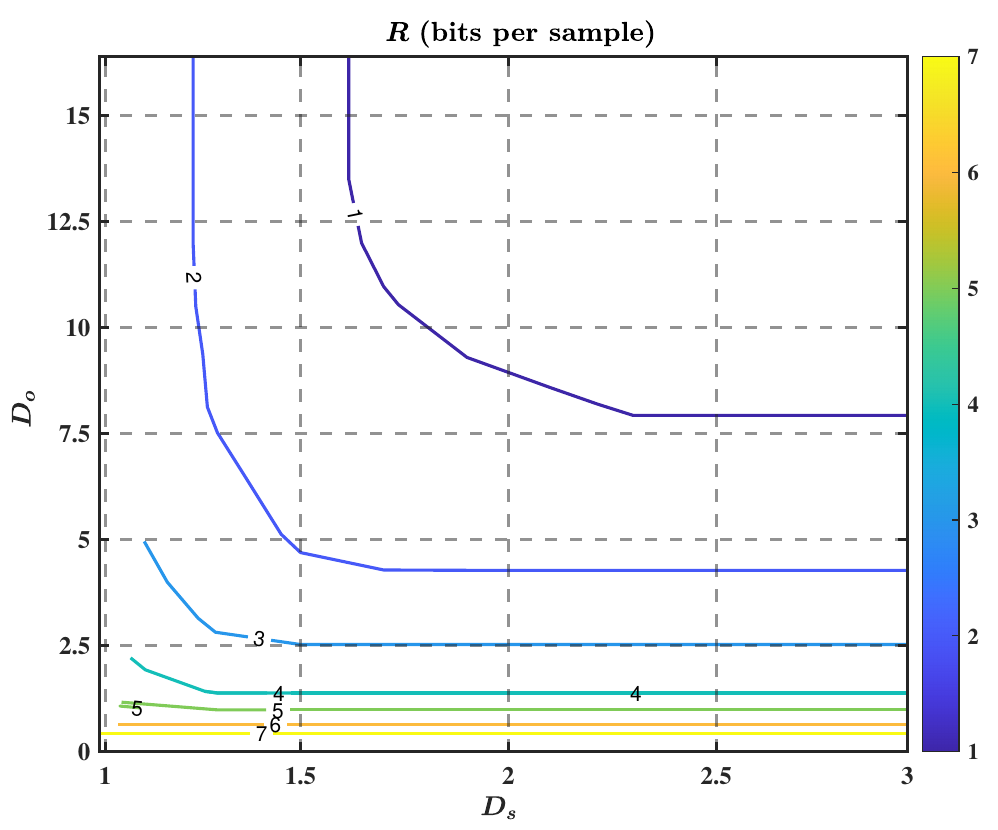}
}
\caption{NESRD  $  \hat{R}_\Theta(D_{\text{o}}, D_{\text{s}}) $ for joint Gaussian source by Algorithm \ref{alg_NESRD}. }
\label{fig_sur_Gaussian}
\end{figure}

Fig. \ref{fig_sur_Gaussian} plots the surface and contour of NESRD  $  \hat{R}_\Theta(D_{\text{o}}, D_{\text{s}}) $ for the joint Gaussian source given in \eqref{para_gaussian}.  
It is easy to see that, with fixed $ D_{\text{s}} $, $  \hat{R}_\Theta(D_{\text{o}}, D_{\text{s}}) $ exhibits a diminishing trend as $ D_{\text{o}} $ increases. However, the decreasing rate gradually attenuates with the augmentation of $ D_{\text{o}} $. Then, when $ D_{\text{o}} $ remains fixed, for smaller values of $ D_{\text{o}} $, $  \hat{R}_\Theta(D_{\text{o}}, D_{\text{s}}) $ is insensitive to variations in $ D_{\text{s}} $; whereas for larger values of $ D_{\text{o}} $, $  \hat{R}_\Theta(D_{\text{o}}, D_{\text{s}}) $ experiences a substantial reduction as $ D_{\text{s}} $ increases. For example, when $ D_{\text{o}} $ is around $ 1 $, changes in $ D_{\text{s}} $ scarcely impact $  \hat{R}_\Theta(D_{\text{o}}, D_{\text{s}}) $. In contrast, when $ D_{\text{o}} $  equals $15$, as $ D_{\text{s}} $ increments, $  \hat{R}_\Theta(D_{\text{o}}, D_{\text{s}}) $ undergoes a notable decrease from around $ 2 $ to $ 0 $.

 \begin{figure}[htbp] 
\centering
\subfigure[ \label{fig_G_Ds} $  \hat{R}_\Theta(D_{\text{o}}, D_{\text{s}}) $  vs. $ R(D_{\text{o}},D_{\text{s}}) $  for fixed $ D_{\text{o}} $ ]{
\includegraphics[width=2.75in]{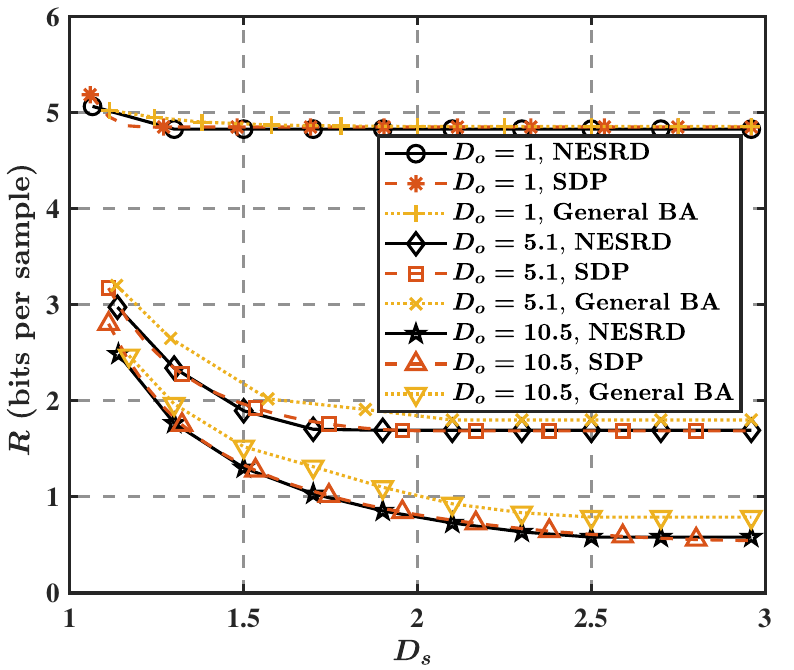}}
 \hspace{0.51in}
\subfigure[ \label{fig_G_Do}  $  \hat{R}_\Theta(D_{\text{o}}, D_{\text{s}}) $  vs. $ R(D_{\text{o}},D_{\text{s}}) $  for fixed $ D_{\text{s}} $]{
\includegraphics[width=2.6in]{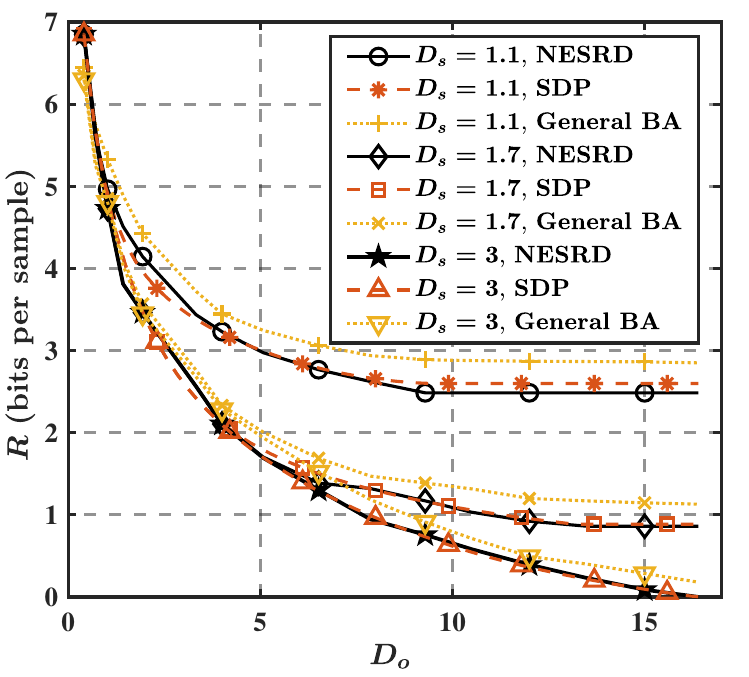}
}
\caption{ Performance comparisons among NESRD,  SDP method\upcite{9844779}, and proposed general BA method for joint Gaussian semantic source.}
\label{fig_SDP_nesrd}
\end{figure}  
 Fig. \ref{fig_SDP_nesrd} compares the performance of NESRD $ \hat{R}_\Theta(D_{\text{o}}, D_{\text{s}}) $, SDP-computed $R(D_{\text{o}}, D_{\text{s}})$\upcite{9844779}, and $R(D_{\text{o}}, D_{\text{s}})$ computed by the proposed  general BA method in Algorthm \ref{alg_ba} for the joint Gaussian source. It is noted that by directly leveraging the distribution of the joint Gaussian source $(X, S)$, the SDP method is able to numerically compute the corresponding SRDF $R(D_{\text{o}}, D_{\text{s}})$ in a stable and efficient manner\upcite{9844779}. Moreover, it shows that $ \hat{R}_\Theta(D_{\text{o}}, D_{\text{s}}) $ is remarkably close to the SDP-computed $ R(D_{\text{o}}, D_{\text{s}}) $ for any distortion couple $ (D_{\text{o}}, D_{\text{s}}) $, which reveals that the proposed NESRD is a good estimator for the SRDF in this example.   However, $ R(D_{\text{o}}, D_{\text{s}}) $ computed by the general BA algorithm exhibits slight deviations from the other two methods at the low rate region ($R \leq 4$ bits per sample). These deviations are potentially attributed to the errors introduced during the discretization of the joint Gaussian source distribution in the general BA algorithm. 
\subsection{Image Datasets}
\subsubsection{Datasets} In our experiment, we calculate the NESRD for MNIST and SVHN datasets by utilizing Algorithm \ref{alg_s_NESRD}.

\subsubsection{Training settings} We consider that each image in the labeled dataset is a sample from extrinsic observation $ P_X $, and its label is a sample from the intrinsic semantic distribution $ P_S $. Besides, we utilize one-hot encoding to represent the labels. Therefore, the semantic sample $ s $ is a one-hot vector with  $10$  dimensions.    The distribution of the latent variable $ Z $ is set to be $ P_Z = \mathcal{N}(0, I_{100}) $ with $ 100 $ being the dimension of the latent space. The generator $ G $  is parameterized by a convolutional neural network (CNN) with  $3$  convolutional layers,  $3$  pooling layers, and $2$ fully connected layers, where the output of the generator $ G $ is used to represent $ \hat{X} $. The classifier $F$  is also parameterized by a CNN with $2$ convolutional layers, $2$ pooling layers, and $2$ fully connected layers. Besides,  the classifier $F$ outputs a probability distribution over 10 classes,   which is used to represent a sample of $ \hat{S} $.  Then, the learning rates of generator $G$ and classifier $ F $ are both set as $ 1\times10^{-4} $ and the training epoch is set as $50$. Besides, we set $ N_1 = 40000 $ for the considered two datasets. Moreover, we adopt the squared-error distortion measures for the extrinsic observation, i.e., $ d_{\text{o}}(x,\hat{x})= \Vert x- \hat{x}\Vert_2^2 $. As for the intrinsic semantic state, its distortion measure $ d_{\text{s}}(s, \hat{s}) $ is defined to be the cross entropy between $ s $ and $ \hat{s} $, i.e.\upcite{pmlr-v97-gao19c},
\begin{equation}
	d_{\text{s}}(s, \hat{s}) = -\sum_{k=0}^{9}s^{(k)}\ln\hat{s}^{(k)}, 
\end{equation}
where $ s = [s^{(0)},\cdots, s^{(9)} ] $ and $ \hat{s} = [\hat{s}^{(0)},\cdots, \hat{s}^{(9)} ] $. Moreover, by the definition of maximum semantic distortion $ D_{\text{max}}^{\text{s}} $ in Remark \ref{Re_property_R}, it is easy to see that $ D_{\text{max}}^{\text{s}} $ is obtained when $ \hat{s} $ is the uniform distribution over the  $10$ classes, i.e., $ \hat{s} = [1/10,\cdots, 1/10] $ and $ D_{\text{max}}^{\text{s}} = - \ln{\frac{1}{10}} \approx 2.3 $.

 \begin{figure}[htbp] 
\centering
\subfigure[\label{fig_3d_mnist} Surface of  $  \hat{R}_\Theta(D_{\text{o}}, D_{\text{s}}) $]{
\includegraphics[width=2.85in]{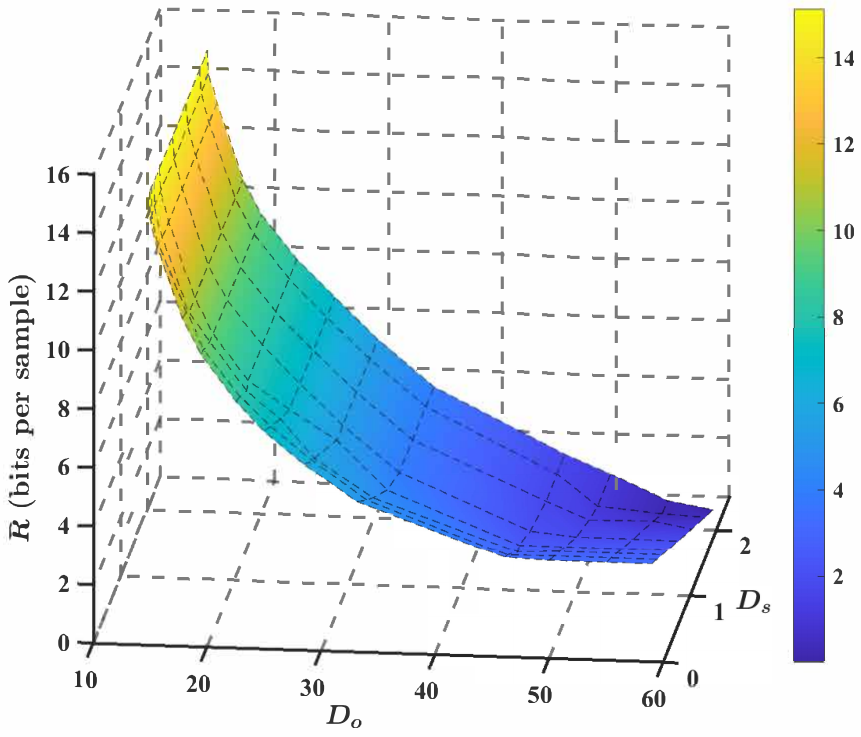}}
 \hspace{0.51in}
\subfigure[ \label{fig_2d_mnist} $  \hat{R}_\Theta(D_{\text{o}}, D_{\text{s}}) $ vs. $ D_{\text{o}} $]{
\includegraphics[width=2.85in]{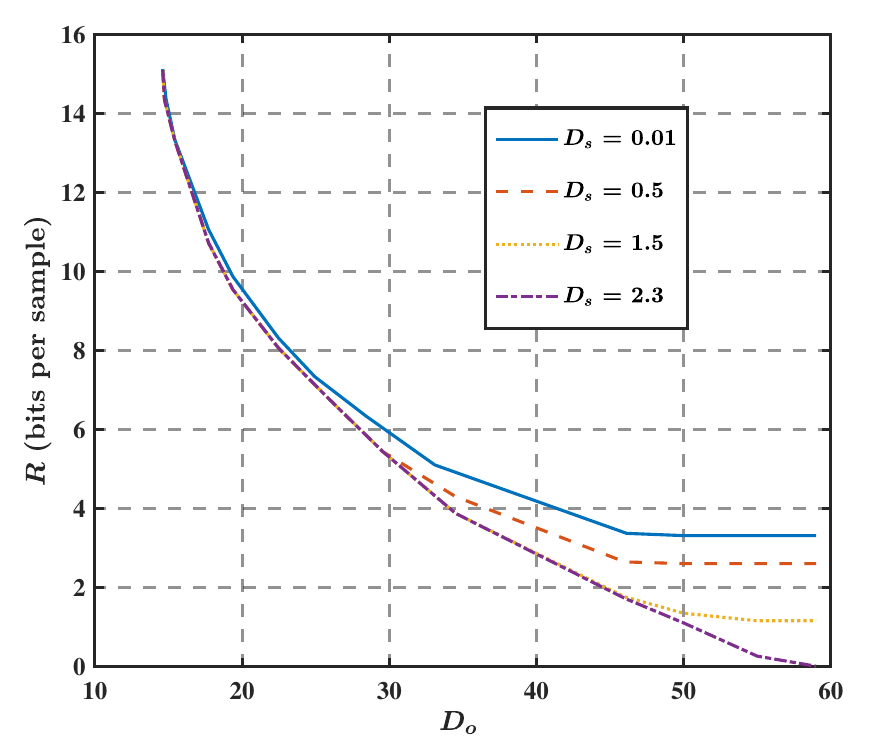}
}
\caption{ NESRD $ \hat{R}_\Theta(D_{\text{o}}, D_{\text{s}}) $ for MNIST dataset.  }
\label{fig_mnist}
\end{figure}  
\subsubsection{Experiments}
For MNIST dataset, we draw the surface of the corresponding NESRD $ \hat{R}_\Theta(D_{\text{o}}, D_{\text{s}}) $, and also plot  $ \hat{R}_\Theta(D_{\text{o}}, D_{\text{s}}) $ as a function of $ D_{\text{o}} $ for fixed values of $ D_{\text{s}} $ in Fig. \ref{fig_mnist}. Specifically, Fig. \ref{fig_3d_mnist} illustrates the overall decreasing trend of $ \hat{R}_\Theta(D_{\text{o}}, D_{\text{s}}) $ with respect to $D_{\text{o}} $ and $ D_{\text{s}} $. Moreover, as depicted in Fig. \ref{fig_2d_mnist}, when extrinsic observation distortion $ D_{\text{o}} $ is less than $ 25 $, semantic distortion $ D_{\text{s}} $ has only a little effect on $ \hat{R}_\Theta(D_{\text{o}}, D_{\text{s}}) $, which indicates that the recovered observation state $\hat{X}$ almost contains the entire semantic information $ S$; Conversely, for $D_{\text{o}} > 25$, the impact of $D_{\text{s}}$ on $\hat{R}_\Theta(D_{\text{o}}, D_{\text{s}}) $ gradually intensifies and the smaller $ D_{\text{s}} $, the larger $\hat{R}_\Theta(D_{\text{o}}, D_{\text{s}}) $. Additionally, for $ D_{\text{s}} =0.01 $ and $D_{\text{o}} >47 $, $\hat{R}_\Theta(D_{\text{o}}, D_{\text{s}}) $ decreases to a constant, which is approximately equal to $ \log10 $. This aligns with the case that only the semantic information, i.e., labels of MNIST images, is nearly losslessly compressed, since the conventional rate-distortion function $ R_{\text{s}}(D_{\text{s}}) $ defined in Remark \ref{Re_property_R} satisfies $R_{\text{s}}(0) = \log10$.
           \begin{figure}[htbp]
	\centering
	\includegraphics[width=4in]{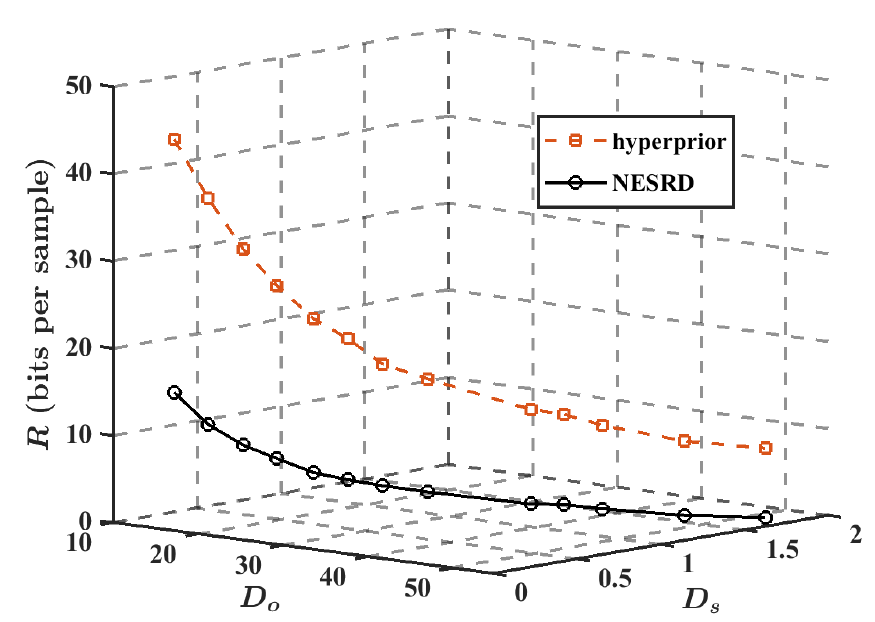}
	\caption{Performance comparisons between NESRD and the hyperprior-based compression method\upcite{balle2018variational} for MNIST dataset.}
	\label{compare_compressAI_mnist} 
    \end{figure}
    
 Then, for MNIST dataset, as shown in Fig. \ref{compare_compressAI_mnist}, we compare NESRD $ \hat{R}_\Theta(D_{\text{o}}, D_{\text{s}}) $ with the hyperprior-based compression method\upcite{balle2018variational} by using the same values of $ D_{\text{o}} $ and $ D_{\text{s}} $. It is easy to see that $ \hat{R}_\Theta(D_{\text{o}}, D_{\text{s}}) $ is significantly smaller than the compression rate of the hyperprior-based compression method. This indicates that there is significant potential for enhancing the performance of current DNN-based compression algorithms when applied to the compression of a specific dataset. For example, most existing compression algorithms typically compress one image at a time. However, jointly compressing multiple images from the same dataset has the potential to further boost the compression performance.

\begin{figure}[htbp]
	\centering
	\includegraphics[width=4in]{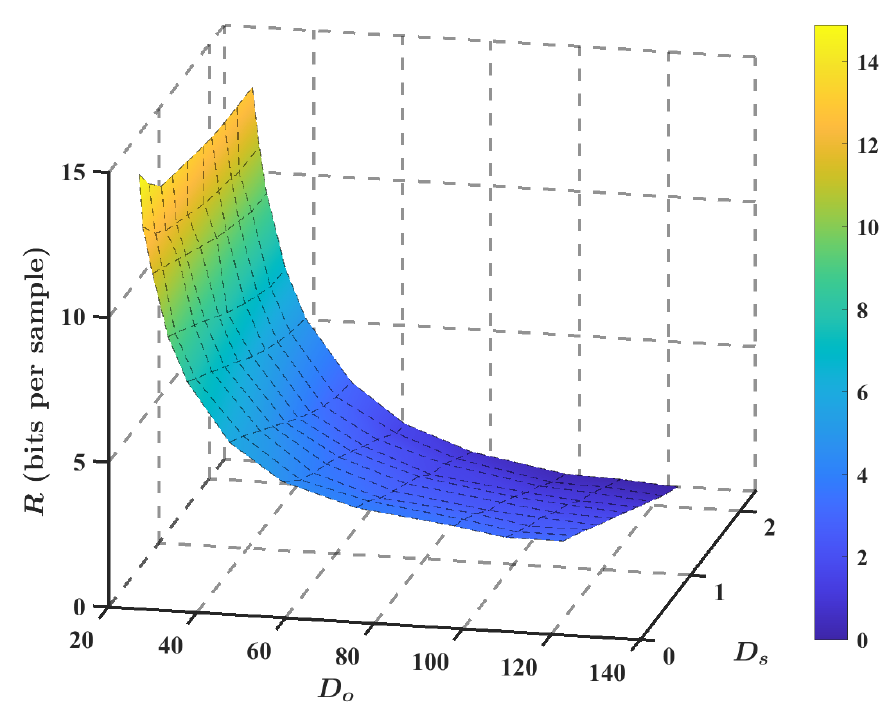}
	\caption{ Surface of  $ \hat{R}_\Theta(D_{\text{o}}, D_{\text{s}}) $ for SVHN dataset.}
	\label{svhn_3d} 
    \end{figure}

\begin{figure}[htbp] 
\centering
\subfigure[ \label{fig_2d_svhn2} $  \hat{R}_\Theta(D_{\text{o}}, D_{\text{s}}) $ vs. $ D_{\text{o}} $]{
\includegraphics[width=2.85in]{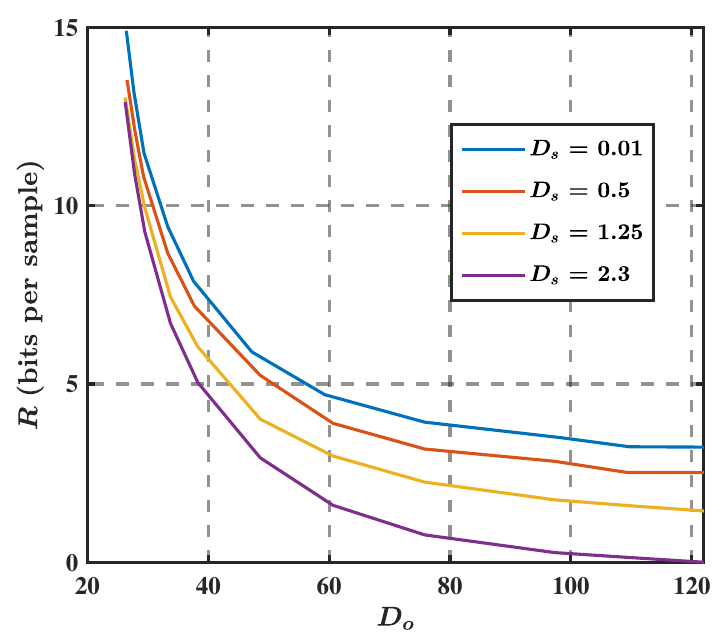}}
 \hspace{0.51in} 
 \subfigure[\label{fig_2d_svhn1}  $  \hat{R}_\Theta(D_{\text{o}}, D_{\text{s}}) $ vs. $ D_{\text{s}} $]{
\includegraphics[width=2.85in]{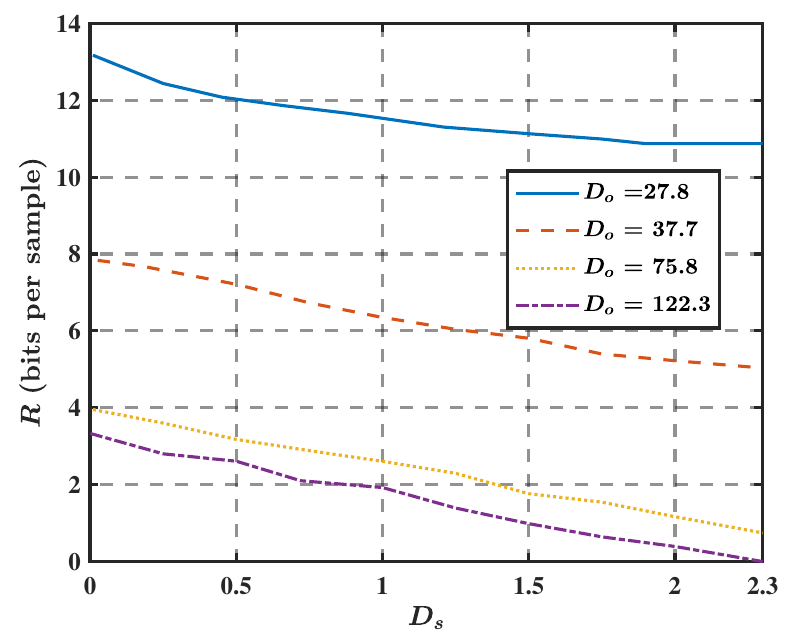}}

\caption{NESRD $ \hat{R}_\Theta(D_{\text{o}}, D_{\text{s}}) $ for SVHN dataset.  }
\label{fig_svhn}
\end{figure}

\begin{figure}[htbp]
	\centering
	\includegraphics[width=4in]{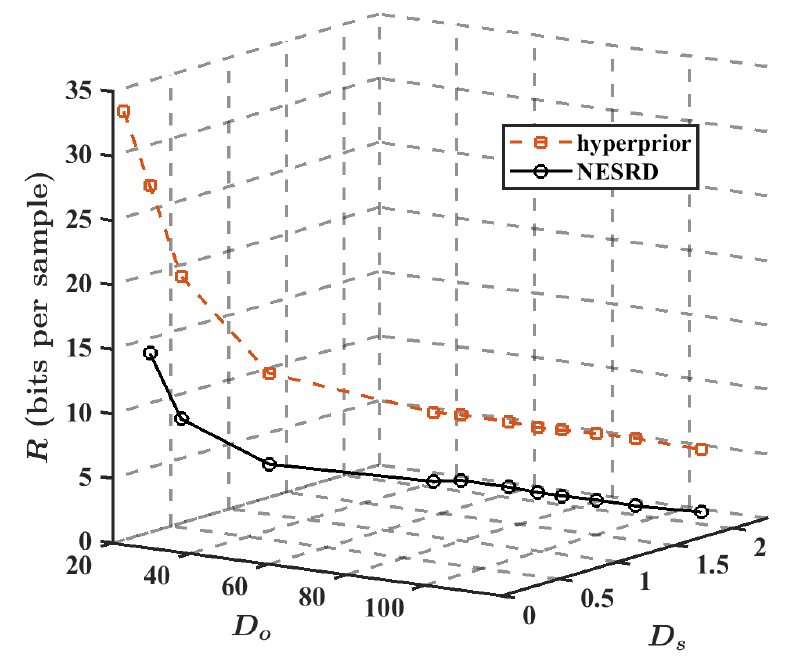}
	\caption{Performance comparisons between NESRD and the hyperprior-based compression method\upcite{balle2018variational} for SVHN dataset.}
	\label{compare_compressAI_svhn} 
    \end{figure}

Moreover, we further show the performance of NESRD for SVHN dataset, which is depicted in Fig. \ref{svhn_3d} and \ref{fig_svhn}. Specifically, similar to Fig. \ref{fig_3d_mnist}, Fig. \ref{svhn_3d} depicts the overall decreasing trend of $ \hat{R}_\Theta(D_{\text{o}}, D_{\text{s}}) $ with respect to $D_{\text{o}} $ and $ D_{\text{s}} $ for SVHN dataset. Notably, the range of $D_{\text{o}} $ for SVHN dataset is quite larger than that of the MNIST dataset, due to the fact that SVHN images are in color, whereas MNIST images are grayscale. Then, in Fig. \ref{fig_2d_svhn2},  when extrinsic distortion $ D_{\text{o}} $ is small, semantic distortion $ D_{\text{s}} $ has only a little effect on $ \hat{R}_\Theta(D_{\text{o}}, D_{\text{s}}) $; as $ D_{\text{o}} $ becomes larger, the impact of $D_{\text{s}}$ on $\hat{R}_\Theta(D_{\text{o}}, D_{\text{s}}) $ gradually intensifies, which performs similar to that of MNIST dataset shown in Fig. \ref{fig_2d_mnist}. Moreover, we  also plot  $ \hat{R}_\Theta(D_{\text{o}}, D_{\text{s}}) $ as a function of $ D_{\text{s}} $ for fixed values of $ D_{\text{o}} $ in Fig. \ref{fig_2d_svhn1}. It is easy to see that variations in $D_{\text{s}}$ can result in changes in $ \hat{R}_\Theta(D_{\text{o}}, D_{\text{s}}) $ up to approximately $\log10$ bits, which aligns with the results of \eqref{ap_R_ineq} since $ R_{\text{s}}(D_{\text{s}}) $ satisfies $R_{\text{s}}(0) = \log10$. Finally,  we also conduct a comparison between NESRD $ \hat{R}_\Theta(D_{\text{o}}, D_{\text{s}}) $ and the hyperprior-based compression method\upcite{balle2018variational} for the SVHN dataset, which is shown in Fig. \ref{compare_compressAI_svhn}. Similar to the findings observed in the MNIST dataset (as depicted in Fig. \ref{compare_compressAI_mnist}), our comparisons reveal that $ \hat{R}_\Theta(D_{\text{o}}, D_{\text{s}}) $ for SVHN is also notably smaller than the compression rate achieved by the hyperprior-based compression method.


\section{Conclusion} 
This paper proposed an SSCC-based framework for point-to-point semantic communications, which explored SRDF to study trad-off among the minimum compression rate, the observation distortion, the semantic distortion, and the channel capacity for generally distributed semantic sources.  Specifically, with the utilization of the generative network, we proposed NESRD, a strongly consistent neural estimator of SRDF for the case with imperfectly known semantic source distribution.  After that, we proposed a general BA algorithm to solve SRDF for the case with perfectly known semantic source distributions and showed the computational complexity of this algorithm. Finally, the experimental results showed the validity of our proposed methods for the joint Gaussian source and some typical image datasets.
\appendices   

\section{Proof of Remark \ref{Re_property_R} }
\label{ap_Re_property_R} 
   First, SRDF $ R(D_{\text{o}}, D_{\text{s}}) $ is bounded as\upcite{9834593}
 \begin{equation}
 	\max\{R_{\text{o}}(D_{\text{o}}),R_{\text{s}}(D_{\text{s}})\} \leq R(D_{\text{o}}, D_{\text{s}})\leq R_{\text{o}}(D_{\text{o}}) + R_{\text{s}}(D_{\text{s}}). \label{ap_R_ineq}
 \end{equation}
 Then, by setting $ D_{\text{o}} = 0 $ in \eqref{ap_R_ineq}, it is easy to see that  $R(0, D_{\text{s}}) \geq \max\{R_{\text{o}}(0),R_{\text{s}}(D_{\text{s}})\} \geq R_{\text{s}}(D_{\text{s}}) $. Moreover, by \eqref{ap_R_ineq}, we have $R(D_{\text{max}}^{\text{o}}, D_{\text{s}})\leq R_{\text{o}}(D_{\text{max}}^{\text{o}}) + R_{\text{s}}(D_{\text{s}}) = R_{\text{s}}(D_{\text{s}}) $ , where the equality holds since $ D_{\text{max}}^{\text{o}} $ is the maximum observation distortion satisfying $ R_{\text{o}}(D_{\text{max}}^{\text{o}}) =0 $\upcite{yeung2008information}.   Finally, considering the monotonic property of $R(D_{\text{o}}, D_{\text{s}})$,  there exists a $ D_{\text{o}}^\prime(D_{\text{s}}) \in [0, D_{\text{max}}^{\text{o}}]  $ such that $ R(D_{\text{o}}, D_{\text{s}}) = R_{\text{s}}(D_{\text{s}}) $  for all $ D_{\text{o}} \geq D_{\text{o}}^\prime(D_{\text{s}}) >0 $. Therefore, we have proved property 2). Besides,  based on the symmetry of $ D_{\text{o}} $ and $ D_{\text{s}} $ in \eqref{def_R}, the proof of property 3) is the same as that of 2). As a result, we have completed this proof.

\section{Proof of Proposition \ref{prop_R1} }
\label{ap_prop_R1}
     First, we present a lemma about $\Lambda_Q(\alpha_1, \alpha_2) $ to reveal its strictly convex property.         \begin{Lemma} 
           \label{a_L_covex}
             $\Lambda_Q(\alpha_1, \alpha_2)$ is strictly convex over the feasible region  $ \{(\alpha_1, \alpha_2):\alpha_1 \leq 0, \alpha_2 \leq 0 \} $.
        \end{Lemma}
        \begin{IEEEproof}
         For simplicity, we define $ \Lambda_Q^{x}(\alpha_1, \alpha_2 )  \triangleq  \ln \textbf{E}_{Q_{(\hat{X}, \hat{S})}} e^{\alpha_1d_{\text{o}}(x, \hat{X}) + \alpha_2\hat{d}_{\text{s}}(x, \hat{S})}  $, and correspondingly, $  \Lambda_Q(\alpha_1, \alpha_2 )  = \textbf{E}_{P_X}[\Lambda_Q^{X}(\alpha_1, \alpha_2 )]   =  \textbf{E}_{P_X}\big[ \ln \textbf{E}_{Q_{(\hat{X}, \hat{S})}} e^{\alpha_1d_{\text{o}}(X, \hat{X}) + \alpha_2\hat{d}_{\text{s}}(X, \hat{S}) } \big] $.
        	 $\Lambda_Q^{x}(\alpha_1, \alpha_2)$ is differentiable with 
        	 \begin{equation}
        	 	\frac{\partial  \Lambda_Q^{x}(\alpha_1, \alpha_2 )}{\partial \alpha_1} = \textbf{E}_{ Q_{(\hat{X}, \hat{S})}}\left( d_{\text{o}}(x, \hat{X}) \frac{e^{\alpha_1d_{\text{o}}(x, \hat{X})+ \alpha_2\hat{d}_{\text{s}}(x, \hat{S})  }}{\textbf{E}_{Q_{(\hat{X}, \hat{S})}}\big[ e^{\alpha_1d_{\text{o}}(x, \hat{X})+ \alpha_2\hat{d}_{\text{s}}(x, \hat{S})  } \big] } \right), \label{ap_part_1}
        	 \end{equation}
        	 and \begin{equation}
        	  \frac{\partial  \Lambda_Q^{x}(\alpha_1, \alpha_2 )}{\partial \alpha_2} = \textbf{E}_{Q_{(\hat{X}, \hat{S})}}\left(  \hat{d}_{\text{s}}(x, \hat{S})\frac{e^{\alpha_1d_{\text{o}}(x, \hat{X})+ \alpha_2\hat{d}_{\text{s}}(x, \hat{S})  }}{\textbf{E}_{Q_{(\hat{X}, \hat{S})}}\big[ e^{\alpha_1d_{\text{o}}(x, \hat{X})+ \alpha_2\hat{d}_{\text{s}}(x, \hat{S})  } \big] } \right).	\label{ap_part_2}
        	 \end{equation}
        	 It can be easily derived that $\frac{\partial  \Lambda_Q^{x}(\alpha_1, \alpha_2 )}{\partial \alpha_1} \geq 0 $ and  $ \frac{\partial  \Lambda_Q^{x}(\alpha_1, \alpha_2 )}{\partial \alpha_2} \geq 0 $ are both held.
        	Moreover, $ \Lambda_Q^{x}(\alpha_1, \alpha_2 ) $ is twice differentiable with 
        	\begin{align}
        		\frac{\partial^2  \Lambda_Q^{x}(\alpha_1, \alpha_2 )}{\partial \alpha_1^2} &= \frac{\textbf{E}_{Q_{(\hat{X}, \hat{S})}} \big[ d_{\text{o}}^2(x, \hat{X})e^{\alpha_1d_{\text{o}}(x, \hat{X})+ \alpha_2\hat{d}_{\text{s}}(x, \hat{S})  } \big] }{\textbf{E}_{Q_{(\hat{X}, \hat{S})}}\big[ e^{\alpha_1d_{\text{o}}(x, \hat{X})+ \alpha_2\hat{d}_{\text{s}}(x, \hat{S})  } \big] }  \notag \\
        		&- \left(   \frac{\textbf{E}_{Q_{(\hat{X}, \hat{S})}} \big[d_{\text{o}}(x, \hat{X})e^{\alpha_1d_{\text{o}}(x, \hat{X})+ \alpha_2\hat{d}_{\text{s}}(x, \hat{S})  }\big]}{\textbf{E}_{Q_{(\hat{X}, \hat{S})}}\big[ e^{\alpha_1d_{\text{o}}(x, \hat{X})+ \alpha_2\hat{d}_{\text{s}}(x, \hat{S})  } \big] } \right)^2,   \label{part_11}
        	\end{align}
        	 \begin{align}
        		\frac{\partial^2  \Lambda_Q^{x}(\alpha_1, \alpha_2 )}{\partial \alpha_2^2} &= \frac{\textbf{E}_{Q_{(\hat{X}, \hat{S})}} \big[ \hat{d}_{\text{s}}^2(x, \hat{S})e^{\alpha_1d_{\text{o}}(x, \hat{X})+ \alpha_2\hat{d}_{\text{s}}(x, \hat{S})  } \big] }{\textbf{E}_{Q_{(\hat{X}, \hat{S})}}\big[ e^{\alpha_1d_{\text{o}}(x, \hat{X})+ \alpha_2\hat{d}_{\text{s}}(x, \hat{S})  } \big] }  \notag \\
        		&- \left(   \frac{\textbf{E}_{Q_{(\hat{X}, \hat{S})}} \big[ \hat{d}_{\text{s}}(x, \hat{S})e^{\alpha_1d_{\text{o}}(x, \hat{X})+ \alpha_2\hat{d}_{\text{s}}(x, \hat{S})  }\big]}{\textbf{E}_{Q_{(\hat{X}, \hat{S})}}\big[ e^{\alpha_1d_{\text{o}}(x, \hat{X})+ \alpha_2\hat{d}_{\text{s}}(x, \hat{S})  } \big] } \right)^2    ,    \label{part_22}
        	\end{align}

and \begin{align}
        		\frac{\partial^2  \Lambda_Q^{x}(\alpha_1, \alpha_2 )}{\partial \alpha_1 \partial\alpha_2}&= \frac{\partial^2  \Lambda_Q^{x}(\alpha_1, \alpha_2 )}{\partial \alpha_2 \partial\alpha_1} = \frac{\textbf{E}_{Q_{(\hat{X}, \hat{S})}} \big[ d_{\text{o}}(x, \hat{X})\hat{d}_{\text{s}} (x, \hat{S})e^{\alpha_1d_{\text{o}}(x, \hat{X})+ \alpha_2\hat{d}_{\text{s}}(x, \hat{S})  } \big] }{\textbf{E}_{Q_{(\hat{X}, \hat{S})}}\big[ e^{\alpha_1d_{\text{o}}(x, \hat{X})+ \alpha_2\hat{d}_{\text{s}}(x, \hat{S})  } \big] } - \notag \\
        		& \left(   \frac{\textbf{E}_{Q_{(\hat{X}, \hat{S})}} \big[ d_{\text{o}}(x, \hat{X})e^{\alpha_1d_{\text{o}}(x, \hat{X})+ \alpha_2\hat{d}_{\text{s}}(x, \hat{S})  }\big]}{\textbf{E}_{Q_{(\hat{X}, \hat{S})}}\big[ e^{\alpha_1d_{\text{o}}(x, \hat{X})+ \alpha_2\hat{d}_{\text{s}}(x, \hat{S})  } \big] } \right)\left(   \frac{\textbf{E}_{Q_{(\hat{X}, \hat{S})}} \big[ \hat{d}_{\text{s}}(x, \hat{S})e^{\alpha_1d_{\text{o}}(x, \hat{X})+ \alpha_2\hat{d}_{\text{s}}(x, \hat{S})  }\big]}{\textbf{E}_{Q_{(\hat{X}, \hat{S})}}\big[ e^{\alpha_1d_{\text{o}}(x, \hat{X})+ \alpha_2\hat{d}_{\text{s}}(x, \hat{S})  } \big] } \right).    \label{part_12}
        	\end{align}
 The above second-order partial derivatives are difficult to be directly analyzed.  For simplicity, we consider a probability measure $ \tilde{Q}^{(x)} $ on $  \mathcal{\hat{X}}\times \mathcal{\hat{S}} $ defined as
\begin{equation}
	\frac{d \tilde{Q}^{(x)}(\hat{x},\hat{s})}{d Q_{(\hat{X},\hat{S})}(\hat{x},\hat{s})} = \frac{e^{\alpha_1d_{\text{o}}(x, \hat{x})+ \alpha_2\hat{d}_{\text{s}}(x, \hat{s}) }}{\textbf{E}_{Q_{(\hat{X}, \hat{S})}}\big[ e^{\alpha_1d_{\text{o}}(x, \hat{X})+ \alpha_2\hat{d}_{\text{s}}(x, \hat{S}) } \big] }. \label{simple_Q}
\end{equation}
Then, combining \eqref{simple_Q} with \eqref{part_11}-\eqref{part_12}, the second-order partial derivatives with respective to $ \Lambda_Q^{x}(\alpha_1, \alpha_2 ) $ are simplified as
\begin{equation}
	\frac{\partial^2  \Lambda_Q^{x}(\alpha_1, \alpha_2 )}{\partial \alpha_1^2}= \textbf{E}_{\tilde{Q}^{(x)}}\left[d_{\text{o}}^2(x, \hat{X})\right] - \left[\textbf{E}_{\tilde{Q}^{(x)}}d_{\text{o}}(x, \hat{X})\right]^2,
	\label{ap_diff_2_1}
\end{equation}
\begin{equation}
	\frac{\partial^2  \Lambda_Q^{x}(\alpha_1, \alpha_2 )}{\partial \alpha_2^2} = \textbf{E}_{\tilde{Q}^{(x)}}\left[\hat{d}_{\text{s}}^2(x, \hat{S})\right] - \left[\textbf{E}_{\tilde{Q}^{(x)}}\hat{d}_{\text{s}}(x, \hat{S})\right]^2,
\end{equation}
and \begin{equation}
	\frac{\partial^2  \Lambda_Q^{x}(\alpha_1, \alpha_2 )}{\partial \alpha_1\alpha_2}= \textbf{E}_{\tilde{Q}^{(x)}}\left[d_{\text{o}}(x, \hat{X})\hat{d}_{\text{s}}(x, \hat{S})\right] - \textbf{E}_{\tilde{Q}^{(x)}}d_{\text{o}}(x, \hat{X})\cdot \textbf{E}_{\tilde{Q}^{(x)}}\hat{d}_{\text{s}}(x, \hat{S}),
\end{equation}
respectively. It shows that $ \frac{\partial^2  \Lambda_Q^{x}(\alpha_1, \alpha_2 )}{\partial \alpha_1^2} $ is  the variance of $ d_{\text{o}}(x, \hat{X}) $ with probability measure  $ \tilde{Q}^{(x)} $, $ \frac{\partial^2  \Lambda_Q^{x}(\alpha_1, \alpha_2 )}{\partial \alpha_2^2} $ is  the variance of $\hat{d}_{\text{s}}(x, \hat{S})  $ with probability measure  $ \tilde{Q}^{(x)} $, and $ \frac{\partial^2  \Lambda_Q^{x}(\alpha_1, \alpha_2 )}{\partial \alpha_1\alpha_2} $ is the covariance of $ (d_{\text{o}}(x, \hat{X}),\hat{d}_{\text{s}}(x, \hat{S})) $ with probability measure  $ \tilde{Q}^{(x)} $. Then, we  have  $  \frac{\partial^2  \Lambda_Q^{x}(\alpha_1, \alpha_2 )}{\partial \alpha_1^2} \geq 0 $, $  \frac{\partial^2  \Lambda_Q^{x}(\alpha_1, \alpha_2 )}{\partial \alpha_2^2} \geq 0 $, and
\begin{equation}
	\frac{\partial^2  \Lambda_Q^{x}(\alpha_1, \alpha_2 )}{\partial \alpha_1^2}\cdot \frac{\partial^2  \Lambda_Q^{x}(\alpha_1, \alpha_2 )}{\partial \alpha_2^2} - \left(\frac{\partial^2  \Lambda_Q^{x}(\alpha_1, \alpha_2 )}{\partial \alpha_1\alpha_2}\right)^2 \geq 0, \label{ap_ineq}
\end{equation}
where \eqref{ap_ineq} is derived by Cauchy-Schwarz inequality.
Therefore, $ \Lambda_Q^{x}(\alpha_1, \alpha_2 ) $ is a convex function with respect to  $ (\alpha_1, \alpha_2 ) $. 
Since $  \Lambda_Q(\alpha_1, \alpha_2 )  = \textbf{E}_{P_X}[\Lambda_Q^{x}(\alpha_1, \alpha_2 )]  $, it can be easily derived that $\Lambda_Q(\alpha_1, \alpha_2 )  $ is also convex.  

Next, we will prove that $\Lambda_Q(\alpha_1, \alpha_2 )$ is actually strictly convex. From \eqref{ap_part_1}, it is easy to verify that 
\begin{equation}
	\lim\limits_{\alpha_1,\alpha_2\to 0} \frac{\partial  \Lambda_Q(\alpha_1, \alpha_2 )}{\partial \alpha_1} =  \textbf{E}_{P_X \times Q_{\hat{X}}}[d_{\text{o}}(X, \hat{X})],
\end{equation} 
and \begin{equation}
	\lim\limits_{\alpha_1,\alpha_2\to -\infty } \frac{\partial  \Lambda_Q(\alpha_1, \alpha_2 )}{\partial \alpha_1} = \textbf{E}_{P_X}[\text{ess inf}_{\hat{X}\sim Q_{\hat{X}}} d_{\text{o}}(X, \hat{X})  ],
\end{equation} 
where $ Q_{\hat{X}} $ is the marginal distribution of $ \hat{X} $  derived from $ Q_{(\hat{X}, \hat{S})} $. Besides, $  \text{ess inf}_{\hat{X}\sim Q_{\hat{X}}}  d_{\text{o}}(x, \hat{X}) $ is 
  the essential infimum of $  d_{\text{o}}(x, \hat{X}) $ of the random variable $ \hat{X} $ with distribution $ Q_{\hat{X}} $, which is defined as $ \text{ess inf}_{\hat{X}\sim Q_{\hat{X}}} d_{\text{o}}(x, \hat{X})= \sup\{t\in\mathbb{R}:Q_{\hat{X}}\{d_{\text{o}}(x, \hat{X})>t\} = 1 \} $. Moreover, considering the case that $ d_{\text{o}}(x, \hat{x}) $ is not essentially constant for all $ x \in \mathcal{X} $, it is easy to see  $ 0\leq \textbf{E}_{P_X}[\text{ess inf}_{\hat{X}\sim Q_{\hat{X}}} d_{\text{o}}(X, \hat{X})  ] <\textbf{E}_{P_X \times Q_{\hat{X}}}[d_{\text{o}}(X, \hat{X})] $\upcite{dembo2002source}. Consequently,  $ \frac{\partial^2\Lambda_Q(\alpha_1, \alpha_2 )}{\partial \alpha_1^2} $ is strictly positive by \eqref{ap_diff_2_1} and the Cauchy-Schwarz inequality.    Similarly, considering the case that $ \hat{d}_{\text{s}}(x, \hat{s}) $ is not essentially constant for all $ x \in \mathcal{X} $, it is easy to see  $ 0\leq \textbf{E}_{P_X}[\text{ess inf}_{\hat{S}\sim Q_{\hat{S}}} \hat{d}_{\text{s}}(X, \hat{S})  ] <\textbf{E}_{P_X \times Q_{\hat{S}}}[\hat{d}_{\text{s}}(X, \hat{S})] $. Then, $ \frac{\partial^2\Lambda_Q(\alpha_1, \alpha_2 )}{\partial \alpha_2^2} $ can also be proved to be strictly positive.

Moreover, together with \eqref{ap_ineq}, $ \Lambda_Q(\alpha_1, \alpha_2 )  $ satisfies
\begin{equation}
	\frac{\partial^2  \Lambda_Q(\alpha_1, \alpha_2 )}{\partial \alpha_1^2}\cdot \frac{\partial^2  \Lambda_Q(\alpha_1, \alpha_2 )}{\partial \alpha_2^2} - \left(\frac{\partial^2  \Lambda_Q(\alpha_1, \alpha_2 )}{\partial \alpha_1\alpha_2}\right)^2 > 0, \label{ap_big}
\end{equation}since $ \hat{d}_{\text{s}}(X, \hat{S}) $ and $ d_{\text{o}}(X, \hat{X}) $ are not linearly dependent. Therefore, we have proved the Hessian matrix $ \nabla^2  \Lambda_Q(\alpha_1, \alpha_2 )\succ 0 $, which means that
  $ \Lambda_Q(\alpha_1, \alpha_2 )  $ is strictly convex in the considered feasible region.	
        	\end{IEEEproof}

Next, to prove this proposition, 
for simplicity, we define 
\begin{equation}
	L_Q(\alpha_1, \alpha_2,D_{\text{o}},D_{\text{s}}) \triangleq \alpha_1D_{\text{o}} + \alpha_2D_{\text{s}} - \Lambda_Q(\alpha_1, \alpha_2 ),  \label{ap_L}
\end{equation}
as the objective function in \eqref{dual_R1}, and correspondingly, $ (\alpha_1^\star, \alpha_2^\star) \triangleq \arg \sup\limits_{\alpha_1, \alpha_2 \leq 0 } L_Q(\alpha_1,\alpha_2,D_{\text{o}},D_{\text{s}})   $.   In the following,  we first derive that $ \sup\limits_{\alpha_1, \alpha_2 \leq 0 } L_Q(\alpha_1, \alpha_2,D_{\text{o}},D_{\text{s}})  $ is an  upper bound of $ R_1(Q_{(\hat{X}, \hat{S})}, D_{\text{o}}, D_{\text{s}}) $. It is easy to show that $ L_Q(\alpha_1, \alpha_2,D_{\text{o}},D_{\text{s}}) $ is smooth and strictly concave with respect to $ \alpha_1 $ and $ \alpha_2 $ by  Lemma \ref{a_L_covex}. Therefore, the optimal point $(\alpha_1^\star, \alpha_2^\star)  $ is unique. Moreover, by the Karush-Kuhn-Tucker (KKT) optimality conditions\upcite{boyd2004},  we have
\begin{equation}
	\Big(D_{\text{o}} -  \frac{\partial\Lambda_Q(\alpha_1, \alpha_2 )}{\partial \alpha_1}\Big) \cdot\alpha_1 =0, \label{ap_kkt_1}
\end{equation}
\begin{equation}
		\Big(D_{\text{s}} -  \frac{\partial\Lambda_Q(\alpha_1, \alpha_2 )}{\partial \alpha_2}\Big) \cdot\alpha_2 =0,\label{ap_kkt_2}
\end{equation}
\begin{equation}
	D_{\text{o}} -  \frac{\partial\Lambda_Q(\alpha_1, \alpha_2 )}{\partial \alpha_1} \geq 0, \quad D_{\text{s}} -  \frac{\partial\Lambda_Q(\alpha_1, \alpha_2 )}{\partial \alpha_2}\geq 0, \quad \alpha_1, \alpha_2 \leq 0.  \label{ap_kkt_3}
\end{equation}
 Then, it is easy to see that there are  four cases of the optimal point $ (\alpha_1^\star, \alpha_2^\star) $ of  $ L_Q(\alpha_1, \alpha_2,D_{\text{o}},D_{\text{s}}) $:
\begin{enumerate}
	\item \label{case_1} $ \alpha_1^\star <0 $ and $ \alpha_2^\star <0 $, which means  $ \frac{\partial\Lambda_Q(\alpha_1, \alpha_2 )}{\partial \alpha_1}|_{(\alpha_1 =\alpha_1^\star,\alpha_2 =\alpha_2^\star )}=D_{\text{o}} $ and $ \frac{\partial  \Lambda_Q(\alpha_1, \alpha_2 )}{\partial \alpha_2}|_{(\alpha_1 =\alpha_1^\star,\alpha_2 =\alpha_2^\star )}=D_{\text{s}} $;
	\item \label{case_2}  $ \alpha_1^\star = 0 $ and $ \alpha_2^\star <0 $, which means $ D_{\text{o}} \geq \frac{\partial  \Lambda_Q(\alpha_1, \alpha_2 )}{\partial \alpha_1}|_{(\alpha_1 =\alpha_1^\star,\alpha_2 =\alpha_2^\star )} $ and $ \frac{\partial  \Lambda_Q(\alpha_1, \alpha_2 )}{\partial \alpha_2}|_{(\alpha_1 =\alpha_1^\star,\alpha_2 =\alpha_2^\star )}=D_{\text{s}} $;
	\item \label{case_3} $ \alpha_1^\star < 0 $ and $ \alpha_2^\star =0 $, which means $    	         D_{\text{s}} \geq \frac{\partial  \Lambda_Q(\alpha_1, \alpha_2 )}{\partial \alpha_2}|_{(\alpha_1 =\alpha_1^\star,\alpha_2 =\alpha_2^\star )} $ and $ \frac{\partial  \Lambda_Q(\alpha_1, \alpha_2 )}{\partial \alpha_1}|_{(\alpha_1 =\alpha_1^\star,\alpha_2 =\alpha_2^\star )}=D_{\text{o}} $;
	\item \label{case_4} $ \alpha_1^\star = 0 $ and $ \alpha_2^\star =0 $, which means $ D_{\text{s}} \geq \textbf{E}_{P_X \times Q_{\hat{X}}}[\hat{d}_{\text{s}}(X, \hat{S})]$ and $ D_{\text{o}} \geq \textbf{E}_{P_X \times Q_{\hat{X}}}[d_{\text{o}}(X, \hat{X})] $.
\end{enumerate}
 For case \ref{case_1}), we define a probability measure $ W $ on $ \mathcal{X} \times \mathcal{\hat{X}}\times \mathcal{\hat{S}} $  as\upcite{dembo2002source}
\begin{equation}
	\frac{d W(x,\hat{x},\hat{s})}{d(P_X \times Q_{(\hat{X},\hat{S})})} = \frac{e^{\alpha_1^\star d_{\text{o}}(x, \hat{x})  + \alpha_2^\star\hat{d}_{\text{s}}(x, \hat{s}) }}{\textbf{E}_{Q_{(\hat{X}, \hat{S})}}\big[ e^{\alpha_1^\star d_{\text{o}}(X, \hat{X})  + \alpha_2^\star\hat{d}_{\text{s}}(X, \hat{S}) } \big] }.
\end{equation}
It is easy to verify that the first marginal distribution  $ W_X $ of $ W $ is $ P_X $,   Then,  we have 
\begin{align}
	R_1(Q_{(\hat{X}, \hat{S})}, D_{\text{o}}, D_{\text{s}}) &\overset{(a)}{\leq} H_{\text{KL}}\big(W|| P_X \times P_{(\hat{X}, \hat{S})} \big) \notag \\ 
	& = \textbf{E}_W \ln \left\{ \frac{e^{\alpha_1^\star d_{\text{o}}(X, \hat{X})  + \alpha_2^\star\hat{d}_{\text{s}}(X, \hat{S}) }}{\textbf{E}_{Q_{(\hat{X}, \hat{S})}}\big[ e^{\alpha_1^\star d_{\text{o}}(X, \hat{X})  + \alpha_2^\star\hat{d}_{\text{s}}(X, \hat{S}) } \big] } \right\}  \notag \\
	&=\alpha_1^\star \textbf{E}_W[d_{\text{o}}(X,\hat{X})] + \alpha_2^\star \textbf{E}_W[\hat{d}_{\text{s}}(X,\hat{S})]  - \textbf{E}_{P_X}\big[ \ln \textbf{E}_{Q_{(\hat{X}, \hat{S})}}  e^{\alpha_1^\star d_{\text{o}}(X, \hat{X})  + \alpha_2^\star\hat{d}_{\text{s}}(X, \hat{S}) } \big] \notag \\
	&\overset{(b)}{=} \alpha_1^\star D_{\text{o}} + \alpha_2^\star D_{\text{s}} - \Lambda_Q(\alpha_1^\star, \alpha_2^\star). \label{R1_less}
\end{align}
where the inequality $(a)$ is derived by the definition of  $R_1(Q_{(\hat{X}, \hat{S})}, D_{\text{o}}, D_{\text{s}}) $ given in \eqref{R1}, and the equality $ (b) $ holds since  $ \textbf{E}_W[d_{\text{o}}(X,\hat{X})]= \frac{\partial  \Lambda_Q(\alpha_1, \alpha_2 )}{\partial \alpha_1}|_{(\alpha_1 =\alpha_1^\star,\alpha_2 =\alpha_2^\star )}=D_{\text{o}} $  and $ \textbf{E}_W[\hat{d}_{\text{s}}(X,\hat{S})]= \frac{\partial\Lambda_Q(\alpha_1, \alpha_2 )}{\partial \alpha_2}|_{(\alpha_1 =\alpha_1^\star,\alpha_2 =\alpha_2^\star )}=D_{\text{s}} $ by \eqref{ap_part_1} and \eqref{ap_part_2}.
Similarly, for case \ref{case_2}) - \ref{case_4}), we can also derive that  $ R_1(Q_{(\hat{X}, \hat{S})}, D_{\text{o}}, D_{\text{s}}) \leq \alpha_1^\star D_{\text{o}} + \alpha_2^\star D_{\text{s}} - \Lambda_Q(\alpha_1^\star, \alpha_2^\star) $.

Finally, by mimicking the proof steps in Section II of Ref. \cite{dembo2002source}, we prove that $ \sup\limits_{\alpha_1, \alpha_2 \leq 0 } L_Q(\alpha_1, \alpha_2,D_{\text{o}},D_{\text{s}}) $ is also a lower bound of $ R_1(Q_{(\hat{X}, \hat{S})}, D_{\text{o}}, D_{\text{s}}) $.   As given in Ref.\cite{dembo2002source},  for any probability measure $ Q^\prime_{(\hat{X}, \hat{S})} $ defined on $ \mathcal{\hat{X}}\times \mathcal{\hat{S}} $ and any measurable function $ \phi: \mathcal{\hat{X}}\times \mathcal{\hat{S}} \to (-\infty, 0] $, the following inequality
\begin{equation}
	H_{\text{KL}}\big(Q^\prime_{(\hat{X}, \hat{S})}|| Q_{(\hat{X}, \hat{S})}\big) \geq \textbf{E}_{Q^\prime_{(\hat{X}, \hat{S})}}\big[\phi(\hat{X}, \hat{S})\big] - \ln \textbf{E}_{Q_{(\hat{X}, \hat{S})}}\big[e^{\phi(\hat{X}, \hat{S})}\big].  \label{inq_measure}
\end{equation}
always holds.
Then, for any  probability measure $ W $  on $ \mathcal{X} \times \mathcal{\hat{X}}\times \mathcal{\hat{S}} $ in \eqref{R1} and any $ x \in \mathcal{X} $, by setting  $ Q^\prime_{(\hat{X}, \hat{S})} = W(\cdot|x) $ and $ \phi(\hat{x}, \hat{s})=\alpha_1^\star d_{\text{o}}(x, \hat{x}) + \alpha_2^\star\hat{d}_{\text{s}}(x,\hat{s}) $, \eqref{inq_measure} can be rewritten as
\begin{align}
	H_{\text{KL}}\big(W(\cdot|x)|| Q_{(\hat{X}, \hat{S})}\big) \geq &\alpha_1^\star\textbf{E}_{W(\cdot|x)}\big[d_{\text{o}}(X, \hat{X})\big] + \alpha_2^\star\textbf{E}_{W(\cdot|x)}\big[\hat{d}_{\text{s}}(X, \hat{S})\big] \notag \\
	 &-\ln \textbf{E}_{Q_{(\hat{X}, \hat{S})}} e^{\alpha_1^\star d_{\text{o}}(x, \hat{X}) + \alpha_2^\star \hat{d}_{\text{s}}(x, \hat{S}) }.
\end{align}
By substituting $x$ with $X$ and taking expectations with respect to $P_X$ on both sides,  we obtain
   \begin{align}
   	H_{\text{KL}}\big(W||P_X \times Q_{(\hat{X}, \hat{S})}\big) &\geq \alpha_1^\star \textbf{E}_{W}\big[d_{\text{o}}(X, \hat{X})\big] + \alpha_2^\star \textbf{E}_{W}\big[\hat{d}_{\text{s}}(X, \hat{S})\big]-  \Lambda_Q(\alpha_1^\star, \alpha_2^\star) \notag \\ 
   	& \overset{(c)}\geq \alpha_1^\star D_{\text{o}} + \alpha_2^\star D_{\text{s}} -  \Lambda_Q(\alpha_1^\star, \alpha_2^\star) \notag \\
   	& = \sup\limits_{\alpha_1, \alpha_2 \leq 0 } L_Q(\alpha_1, \alpha_2,D_{\text{o}},D_{\text{s}}),
   \end{align}  
   where the inequality $(c)$ holds since $\alpha_1^\star, \alpha_2^\star \leq 0$,  and $\textbf{E}_{W}\big[d_{\text{o}}(X, \hat{X})\big] \leq D_{\text{o}}$ and  $\textbf{E}_{W}\big[\hat{d}_{\text{s}}(X, \hat{S})\big] \leq D_{\text{s}}$  in \eqref{R1}.
   Moreover, since $ W $ is  chosen arbitrarily, we can derive that  $ R_1\big(Q_{(\hat{X}, \hat{S})}, D_{\text{o}}, D_{\text{s}}\big)  \geq \sup\limits_{\alpha_1, \alpha_2 \leq 0 } L_Q(\alpha_1, \alpha_2,D_{\text{o}},D_{\text{s}}) $. Thus, combining this with inequality \eqref{R1_less} completes the proof.  
  
   \section{Proof of Proposition \ref{Prop_R_R1} }
\label{ap_Prop_R_R1}
    	From the definitions of $ R(D_{\text{o}}, D_{\text{s}}) $ and $R_1(Q_{(\hat{X}, \hat{S})}, D_{\text{o}}, D_{\text{s}})$ in \eqref{def_R} and \eqref{R1}, it is easy to obtain $  \inf\limits_{Q_{(\hat{X}, \hat{S})}} R_1(Q_{(\hat{X}, \hat{S})}, D_{\text{o}}, D_{\text{s}}) \leq R(D_{\text{o}}, D_{\text{s}}) $.  Moreover, based on the definition of KL distance, \eqref{R1} can be equivalently written as 
    	\begin{align}
    		R_1\big(Q_{(\hat{X}, \hat{S})}, D_{\text{o}}, D_{\text{s}}\big) &=  \min\limits_{ \substack{ P_{(\hat{X}, \hat{S})|X} \\ \textbf{E}\left[\hat{d}_{\text{s}}(X, \hat{S})\right] \leq D_{\text{s}}  \\  \textbf{E}\left[d_{\text{o}}(X, \hat{X})\right] \leq D_{\text{o}} }}\left\{ H_{\text{KL}}\Big(P_{(X, \hat{X}, \hat{S})}|| P_X \times P_{(\hat{X}, \hat{S})} \Big) + H_{\text{KL}}(P_{(\hat{X}, \hat{S})}||Q_{(\hat{X}, \hat{S})} )  \right\},  \label{ap_R1_2} 
    	\end{align}
    	which implies that $ R_1\big(Q_{(\hat{X}, \hat{S})}, D_{\text{o}}, D_{\text{s}}\big) \geq R(D_{\text{o}}, D_{\text{s}}) $ holds  for any fixed distribution $ Q_{(\hat{X}, \hat{S})} $. Therefore, \eqref{R_Q} is derived. 
    	
    	 Next, denoting $ P_{(\hat{X}, \hat{S})}^\star   $ as the optimal  distribution of $(\hat{X}, \hat{S})$ in problem \eqref{def_R}, we aim to prove that the optimal distribution of $ (\hat{X}, \hat{S})$ in problem \eqref{R_Q} satisfies $Q_{(\hat{X}, \hat{S})}^\star = P_{(\hat{X}, \hat{S})}^\star $. On the one hand, if $Q_{(\hat{X}, \hat{S})}^\star = P_{(\hat{X}, \hat{S})}^\star $, it is easy to obtain $ R_1\big(Q_{(\hat{X}, \hat{S})}^\star, D_{\text{o}}, D_{\text{s}}\big) = R(D_{\text{o}},D_{\text{s}}) $ from \eqref{ap_R1_2}; on the other hand, if $Q_{(\hat{X}, \hat{S})}^\star \neq P_{(\hat{X}, \hat{S})}^\star $, we can show that $ R_1\big(Q_{(\hat{X}, \hat{S})}^\star, D_{\text{o}}, D_{\text{s}}\big) > R(D_{\text{o}},D_{\text{s}}) $ always holds in the following. Specifically, by \eqref{ap_R1_2},  we have    
    	\begin{align}
    		R_1\big(Q_{(\hat{X}, \hat{S})}^\star, D_{\text{o}}, D_{\text{s}}\big) &=  \min\limits_{ \substack{ P_{(\hat{X}, \hat{S})|X} \\ \textbf{E}\left[\hat{d}_{\text{s}}(X, \hat{S})\right] \leq D_{\text{s}}  \\  \textbf{E}\left[d_{\text{o}}(X, \hat{X})\right] \leq D_{\text{o}} }}\left\{ H_{\text{KL}}\Big(P_{(X, \hat{X}, \hat{S})}|| P_X \times P_{(\hat{X}, \hat{S})} \Big) + H_{\text{KL}}(P_{(\hat{X}, \hat{S})}||Q_{(\hat{X}, \hat{S})}^\star )  \right\} \label{ap_Q_star_prob} \\ &= H_{\text{KL}}\Big(P_{(X, \hat{X}, \hat{S})}^\prime|| P_X \times P_{(\hat{X}, \hat{S})}^\prime \Big) + H_{\text{KL}}(P_{(\hat{X}, \hat{S})}^\prime||Q_{(\hat{X}, \hat{S})}^\star ) , \label{ap_R1_prime}
     	\end{align}
    	where $ P_{(X, \hat{X}, \hat{S})}^\prime $ is the optimal distribution of $ (X, \hat{X}, \hat{S}) $ in problem \eqref{ap_Q_star_prob},  and  $ P_{(\hat{X}, \hat{S})}^\prime $ is the corresponding marginal distribution obtained from $ P_{(X, \hat{X}, \hat{S})}^\prime $.  Then, there are two cases to study $ R_1\big(Q_{(\hat{X}, \hat{S})}^\star, D_{\text{o}}, D_{\text{s}}\big) $:
    	\begin{enumerate}
    		\item  if $ P_{(\hat{X}, \hat{S})}^\prime =  P_{(\hat{X}, \hat{S})}^\star$, $ R_1\big(Q_{(\hat{X}, \hat{S})}^\star, D_{\text{o}}, D_{\text{s}}\big) $ satisfies
    	\begin{align}
    			R_1\big(Q_{(\hat{X}, \hat{S})}^\star, D_{\text{o}}, D_{\text{s}}\big) &\geq R(D_{\text{o}},D_{\text{s}})+ H_{\text{KL}}(P_{(\hat{X}, \hat{S})}^\star||Q_{(\hat{X}, \hat{S})}^\star ) \label{ap_R1_big1} \\ &> R(D_{\text{o}},D_{\text{s}}) , \label{ap_R1_big2}
    	\end{align}
    	where inequality \eqref{ap_R1_big1} is derived by \eqref{def_R} and $ P_{(\hat{X}, \hat{S})}^\prime =  P_{(\hat{X}, \hat{S})}^\star$, and   \eqref{ap_R1_big2} is obtained since  $Q_{(\hat{X}, \hat{S})}^\star \neq P_{(\hat{X}, \hat{S})}^\star $.
    	\item if $ P_{(\hat{X}, \hat{S})}^\prime \neq  P_{(\hat{X}, \hat{S})}^\star$, \eqref{ap_R1_prime} implies 
    	\begin{align}
    		R_1\big(Q_{(\hat{X}, \hat{S})}^\star, D_{\text{o}}, D_{\text{s}}\big)&>R(D_{\text{o}},D_{\text{s}})+H_{\text{KL}}(P_{(\hat{X}, \hat{S})}^\prime||Q_{(\hat{X}, \hat{S})}^\star ) \\&\geq R(D_{\text{o}},D_{\text{s}}).    	\end{align}
    	\end{enumerate}
    		Therefore, we have proved  $ R_1\big(Q_{(\hat{X}, \hat{S})}^\star, D_{\text{o}}, D_{\text{s}}\big) > R(D_{\text{o}},D_{\text{s}}) $ if  $Q_{(\hat{X}, \hat{S})}^\star \neq P_{(\hat{X}, \hat{S})}^\star $. In conclusion, we have completed this proof.
   
\section{Proof of Proposition \ref{prop_Q} }
   \label{ap_prop_Q}
   	 For any fixed $ \alpha_1 $ and $ \alpha_2 $, $ \alpha_1, \alpha_2 \leq 0 $, and given $  D_{\text{o}}^{\star}$, $D_{\text{s}}^{\star} $, and $ Q^{\star}_{(\hat{X}, \hat{S})} $ in Proposition \ref{prop_Q}, $ L_{Q^{\star}}(\alpha_1, \alpha_2, D_{\text{o}}^{\star}, D_{\text{s}}^{\star})  $ defined in \eqref{ap_L} can be expressed as 
   	\begin{align}
   	L_{Q^{\star}}(\alpha_1, \alpha_2, D_{\text{o}}^{\star}, D_{\text{s}}^{\star}) &  = \alpha_1D_{\text{o}}^{\star}+ \alpha_2D_{\text{s}}^{\star}  - \Lambda_{Q^{\star}}(\alpha_1, \alpha_2) \notag \\
   		&=   \inf\limits_{Q_{(\hat{X}, \hat{S})}} \alpha_1D_{\text{o}}^{\star}+ \alpha_2D_{\text{s}}^{\star}  - \Lambda_{Q}(\alpha_1, \alpha_2) \label{ap_L_1} \\
   		&\leq \inf\limits_{Q_{(\hat{X}, \hat{S})}}\sup\limits_{\hat{\alpha}_1, \hat{\alpha}_2 \leq 0 }\hat{\alpha}_1 D_{\text{o}}^{\star} + \hat{\alpha}_2 D_{\text{s}}^{\star} - \Lambda_Q(\hat{\alpha}_1, \hat{\alpha}_2) \label{ap_L_2} \\
   		&  = R(D_{\text{o}}^{\star}, D_{\text{s}}^{\star}) \label{ap_low_R},
   	\end{align}
   	where \eqref{ap_L_1} is obtained by \eqref{pro_Lambda}, \eqref{ap_L_2} holds since its inner supremum is always larger than or equal to $ \alpha_1D_{\text{o}}^{\star}+ \alpha_2D_{\text{s}}^{\star}  - \Lambda_{Q}(\alpha_1, \alpha_2) $, and \eqref{ap_low_R} is given by \eqref{exp_R}. Moreover, by \eqref{ap_low_R}, $ R(D_{\text{o}}^{\star}, D_{\text{s}}^{\star}) $ satisfies
   	\begin{align}
   		R(D_{\text{o}}^{\star}, D_{\text{s}}^{\star}) &= \inf\limits_{Q_{(\hat{X}, \hat{S})}}\sup\limits_{\hat{\alpha}_1, \hat{\alpha}_2 \leq 0 }\hat{\alpha}_1D_{\text{o}}^{\star}  + \hat{\alpha}_2D_{\text{s}}^{\star} - \Lambda_Q(\hat{\alpha}_1, \hat{\alpha}_2) \notag \\
   		&\leq \sup\limits_{\hat{\alpha}_1, \hat{\alpha}_2 \leq 0 }\hat{\alpha}_1 D_{\text{o}}^{\star} + \hat{\alpha}_2D_{\text{s}}^{\star} - \Lambda_{Q^{\star}}(\hat{\alpha}_1, \hat{\alpha}_2) \label{ap_mid_sup} \\
   		& = 	L_{Q^{\star}}(\alpha_1, \alpha_2, D_{\text{o}}^{\star}, D_{\text{s}}^{\star}), \label{ap_up_R}
   	\end{align}
   	where \eqref{ap_up_R} holds since $ (\alpha_1, \alpha_2) $ is the stationary point of the objective function in \eqref{ap_mid_sup} given in \eqref{1_part_1} and \eqref{1_part_2}.   Finally,
   	Combining \eqref{ap_low_R} with \eqref{ap_up_R}, we have
   	\begin{align}
   		R(D_{\text{o}}^{\star}, D_{\text{s}}^{\star}) = L_{Q^{\star}}(\alpha_1, \alpha_2, D_{\text{o}}^{\star}, D_{\text{s}}^{\star}).\label{ap_compute_R}
   	\end{align}
 Together with \eqref{ap_L}, \eqref{R_star} is obtained and we have completed this proof.

\end{document}